\documentclass[journal]{IEEEtran}

\usepackage{amsmath}
\usepackage[center]{subfigure}
\usepackage{lineno,hyperref}
\usepackage{xcolor}
\usepackage{extarrows}
\usepackage{algorithm}
\usepackage{algorithmic}
\usepackage{setspace}
\usepackage{amssymb}
\usepackage{bm}
\usepackage{graphicx}
\usepackage[justification=centering]{caption}
\usepackage{cite}
\usepackage{subfigure}
\usepackage{url}
\usepackage{multirow}
\usepackage{tabularx}
\usepackage{booktabs}
\usepackage{tikz}
\usetikzlibrary{arrows,positioning,fit,shapes,calc,decorations.pathreplacing,matrix,patterns}
\newcommand{\rev}{\textcolor{black}}

\newcommand{\larrow}{\overset{\scriptscriptstyle\leftarrow}}
\newcommand{\rarrow}{\overset{\scriptscriptstyle\rightarrow}}

\newcommand{\ba}{\boldsymbol{a}}
\newcommand{\bb}{\boldsymbol{b}}

\newcommand{\bx}{\boldsymbol{x}}

\newcommand{\bq}{\boldsymbol{q}}

\newcommand{\mA}{\mathcal{A}}

\begin{document}
\title{Message Passing Based Structured Sparse Signal Recovery for Estimation of OTFS Channels with Fractional Doppler Shifts}
\author{ Fei Liu, Zhengdao Yuan, Qinghua Guo, \IEEEmembership{Senior Member, IEEE}, Zhongyong Wang and Peng Sun
\thanks{The work of F. Liu, Z. Yuan and Z. Wang was supported by National Natural Science Foundation of China (61901417,U1804152) and Science and Technology Research Project of Henan Province (212102210556, 202102210313, 212102210566).
Corresponding authors: Qinghua Guo and Zhongyong Wang.}
\thanks{F. Liu, Z. Wang and P. Sun are with the School of Information Engineering, Zhengzhou University, Zhengzhou 450002, China, (e-mail: ieliufei@hotmail.com, zywangzzu@gmail.com, iepengsun@zzu.edu.cn)}
\thanks{Z. Yuan is with the Artificial Intelligence Technology Engineering Research Center, Open University of Henan, and also with the School of Information Engineering, Zhengzhou University, Zhengzhou 450002, China (e-mail: yuan\_zhengdao@163.com).}
\thanks{Q. Guo is with the School of Electrical, Computer and Telecommunications Engineering, University of Wollongong, Wollongong, NSW 2522, Australia  (e-mail: qguo@uow.edu.au).}
}
\maketitle

\begin{abstract}
The orthogonal time frequency space (OTFS) modulation
has emerged as a promising modulation scheme for high mobility wireless communications. To enable efficient OTFS detection in the delay-Doppler (DD) domain, the DD domain channels need to be acquired accurately. To achieve the low latency requirement in future wireless communications, the time duration of the OTFS block should be small, therefore fractional Doppler shifts have to be considered to avoid significant modelling errors due to the assumption of integer Doppler shifts. However there lack investigations on the estimation of OTFS channels with fractional Doppler shifts in the literature. In this work, we develop a channel estimator for OTFS with particular attention to fractional Doppler shifts, and both bi-orthogonal waveform and rectangular waveform are considered. Instead of estimating the DD domain channel directly, we estimate the channel gains and (fractional) Doppler shifts that parameterize the DD domain channel. The estimation is formulated as a structured sparse signal recovery problem with a Bayesian treatment. Based on a factor graph representation of the problem, an efficient message passing algorithm is developed to recover the structured sparse signal (thereby the OTFS channel). The Cramer-Rao Lower Bound (CRLB) for the estimation is developed and the effectiveness of the algorithm is demonstrated through simulations.
\end{abstract}


\section{Introduction}
\IEEEPARstart {R}{encently}, the orthogonal time frequency space (OTFS) modulation has been proposed to achieve reliable communications in high-mobility scenarios \cite{Hadani2017},\cite{Raviteja2018},\cite{OAP}. OTFS provides both time and frequency diversity because each symbol is spread over the time and frequency domains through the two dimensional inverse symplectic finite Fourier transform (ISFFT) \cite{Hadani2017},\cite{Raviteja2018}. When the number of channel paths is small, the effective channel in the delay-Doppler (DD) domain is sparse, which allows efficient data detection using the message passing techniques \cite{Raviteja2018}. A variety of OTFS detection methods have been proposed in the literature to harvest the time and frequency diversity promised by OTFS \cite{2020Iterative,  Raviteja2019, OTFSHighDopplerChannel, performance2020, LiA2017, Raviteja2019Practical, zemen2017, otfslmmserecv}. However, all the detection methods assume perfect channel state information, which has to be estimated in practice.

In OTFS, the delay shifts and Doppler shifts are discretized in the DD domain. In general, a wideband system is able to provide sufficient delay resolution, so that fractional delay shifts do not need to be considered \cite{fundWC}. However, the Doppler resolution depends on the time duration of the OTFS block. To fulfill the low latency requirement in future wireless communications, the time duration of the OTFS block should be relatively small, hence fractional Doppler shifts have to be considered to avoid significant modelling errors due to the assumption of integer Doppler shifts \cite{Raviteja2018},\cite{pilotref}. Therefore factional Doppler shifts have to be considered in OTFS channel estimation.

A number of channel estimation methods have been proposed in the literature. In \cite{MIMOOTFSDETECTIONEST}, an entire OTFS block is used to accommodate pilot symbols for channel estimation, and the estimated channel is used for data detection in the subsequent OTFS block. This results in a considerable loss in  spectrum efficiency and the detection performance may deteriorate due to the channel variation between two OTFS blocks.
To solve this problem, pilot and data symbols are placed in the same OTFS block in \cite{pilotref}, where guard interval is used to avoid interference between pilot and data symbols. With this scheme, channel estimation and data detection can be performed for the same OTFS block. In \cite{chestmassivemimo}, channel estimation in massive multiple input and multiple output (MIMO)-OTFS  systems is studied, where the downlink time-varying massive MIMO channels are transformed to the delay-Doppler-angle domain, and the channel estimation is formulated as a sparse signal recovery problem. In \cite{uplinkotfsmassivemimo}, the uplink-aided downlink massive MIMO-OTFS channel estimation over the delay-Doppler-angle domain is studied. However, in these works, only integer Doppler shifts are considered, except \cite{pilotref}. In \cite{pilotref}, a single pilot symbol is used to facilitate the design of a low complexity thresholding method \rev{to acquire the channel in the DD domain}. Although factional Doppler shifts are considered for OTFS with the bi-orthogonal waveform in \cite{pilotref}, it is not clear how to estimate the channel with fractional Doppler shifts for OTFS with a more practical waveform, e.g., the rectangular waveform. In addition, the method estimates the DD domain channel directly, and does not provide the estimates of the channel gains and fractional Doppler shifts.

In this work, we address the issue of OTFS channel estimation with particular attention to fractional Doppler shifts. As the OTFS channel is parameterized by the channel gains and fractional Doppler shifts, we estimate these parameters rather than the DD domain channel directly, mainly due to two reasons. Firstly, this can lead to much better performance as the number of variables to be estimated is significantly smaller. Secondly, the estimates of Doppler shifts can be useful, e.g., for estimating the velocity of mobile users.  In addition, both the bi-orthogonal waveform and the rectangular waveform are considered. We formulate the \rev{OTFS} channel estimation as a structured signal recovery problem, which is solved using Bayesian inference. With a factor graph representation of the problem, a message passing algorithm is developed to estimate the channel gains and fractional Doppler shifts jointly. In contrast to \cite{pilotref}, we show that using a few pilot symbols rather than a single pilot symbol \rev{is} more preferable as the peak to average power ratio (PAPR) of OTFS signals can be significantly reduced. Our proposed algorithm can work with either single pilot symbol or multiple pilot symbols, and significantly outperforms the state-of-the-art method. The Cramer-Rao lower bound (CRLB) for the \rev{channel} estimation is derived to verify the performance of the proposed algorithm. The bit error rates (BER) of the OTFS system with perfect channel and estimated channel are also compared to demonstrate the effectiveness of the proposed algorithm.

\rev{The main contributions of this work are summarized as follows:}
\begin{itemize}
	\item \rev{The OTFS DD domain channel is acquired by estimating the relevant parameters, i.e., the channel gains and fractional Doppler shifts. The estimation of the parameters is formulated as a novel structured sparse signal recovery problem with a Bayesian treatment.}
	\item \rev{Both the bi-orthogonal waveform and the rectangular waveform are considered in this work. To the best of our knowledge, the investigations on the fractional Doppler shift estimation are very limited in the literature. The work in \cite{pilotref}, where the DD domain channel is estimated directly, does not provide fractional Doppler estimate explicitly, and it is not applicable to the case of the rectangular waveform.}
	\item \rev{ Due to the uniqueness of the formulated structured sparse signal recovery problem, a dedicated message passing algorithm is proposed to efficiently solve the problem. The algorithm is able to work with either a single pilot symbol or multiple pilot symbols, which can significantly reduce the PAPR of the OTFS signal.}
	\item \rev{To evaluate the performance of the estimator, the CRLB is derived. Comparison results with existing methods are provided to demonstrate the advantages of the proposed method.} 	
\end{itemize}
 For simplicity, this work is focused on a plain OTFS system, but the proposed algorithm can be extended for a more complex system, such as MIMO-OTFS.

The remainder of the paper is organized as follows. In Section \ref{sec:model}, we introduce OTFS modulation and demodulation, and OTFS input-output relation in the DD domain. 
\rev{Then} we formulate the channel estimation as a structured sparse signal recovery problem in Section \ref{sec:formulation}, and develop the message passing based algorithm in Section \ref{sec:algorithm}. The CLRB for channel estimation is derived in Section \ref{sec:crlb}. Simulation results are provided in Section \ref{sec:simulation}, followed by conclusions in Section \ref{sec:conclusion}.

\textit{Notations}- Boldface lower-case and upper-case letters denote column vectors and matrices, respectively. The superscripts $(\cdot)^T$ and $(\cdot)^*$ represent transpose and conjugate operations, respectively. We \rev{use} $[\cdot]_M$ to denote the modulo-$M$ operation. The probability density function of a complex Gaussian variable with mean {$\hat x$} and variance $\nu_x$ is represented by $\mathcal{CN}(x;{\hat x},\nu_x)$. The Gamma distribution with scale $\epsilon$ and rate $\eta$ is represented as $Ga(x; \epsilon, \eta)$. The uniform distribution over the range $[a, b]$ is represented by $U[a,b]$. The relation $f(x)=cg(x)$ for some positive constant $c$ is written as $f(x)\propto g(x)$. The notation $\otimes$ represents the Kronecker product, and $\ba\cdot\bb$ and $\ba\cdot/\bb$ represent the element-wise product and division between vectors $\ba$ and $\bb$, respectively. We use $Tr(\boldsymbol{A})$ \rev{and} $\boldsymbol{I}_N$ to denote the trace of $\boldsymbol{A}$ and \rev{an} $N\times N$ identity matrix, respectively.
We use $|x|^2$ to denote the magnitude squared operation for $x$ and use $||\bx||_2^2$ to denote the squared norm of vector $\bx$. We use $q_j$ to denoted the $j$th element of $\bq$. The superscript $t$ of $\textbf{s}^t$ denotes the iteration index in an iterative algorithm. The notation $\mathbb{E}(x)$ is used to denote the expectation of the random variable $x$. \rev{The notations $\mathcal{R}[a]$ and $\mathcal{I}[a]$ denote the real and imaginary parts of the complex variable $a$, respectively}.

\tikzstyle{block} = [draw, fill=white, rectangle, 
minimum height=2.5em, minimum width=4em, align = center]
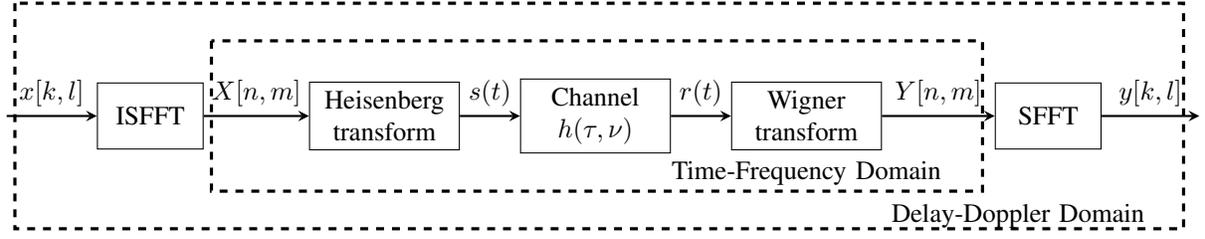
\begin{figure*}[htbp]
	\centering
	\begin{tikzpicture}[auto, node distance=1.7cm,>=latex']
		
	\node[block, ](isfft){ISFFT};
	\node[block, right = 1.4cm of isfft, text width=5em](heisenberg){Heisenberg \\ transform};
	\node[block, right = 0.8cm of heisenberg, text width=5em](channel){Channel \\ $h(\tau,\nu)$};
	\node[block, right = 0.8cm of channel, text width=5em](wigner){Wigner \\ transform};
	\node[block, right = 1.5cm of wigner](sfft){SFFT};
	
	\node[below = 0.02 of wigner, ](tf){Time-Frequency Domain};
	\node[below = 0.6cm of sfft, xshift=-0.4cm](dd){Delay-Doppler Domain};
	
	\node[left = 1.2cm of isfft](xkl){};
	\draw[thick, ->, >=stealth] (xkl) -- node[above]{$x[k,l]$}(isfft);
	\draw[thick, ->, >=stealth] (isfft) -- node[above]{$X[n,m]$}(heisenberg);
	\draw[thick, ->, >=stealth] (heisenberg) -- node[above]{$s(t)$}(channel);
	\draw[thick, ->, >=stealth] (channel) -- node[above]{$r(t)$}(wigner);
	\draw[thick, ->, >=stealth] (wigner) -- node[above]{$Y[n,m]$}(sfft);
	\node[right = 1.3cm of sfft](ykl){};
	\draw[thick, ->, >=stealth] (sfft) -- node[above]{$y[k,l]$}(ykl);
	
	\draw[very thick, dashed] ($(heisenberg)+(-2.3,1)$) rectangle ($(wigner)+(2.34,-1)$);
	\draw[very thick, dashed] ($(isfft)+(-1.8,1.5)$) rectangle ($(sfft)+(1.8,-1.5)$);
	
	\end{tikzpicture}
\caption{OTFS modulation and demodulation \cite{Raviteja2018}. }
\label{fig:model}
\end{figure*}

\section{OTFS System Model} \label{sec:model} 

\subsection{System Model in the DD Domain}
As shown in Fig. \ref{fig:model}, the OTFS modulation and demodulation are   implemented with 2D inverse SFFT (ISFFT) and SFFT at the transmitter and receiver, respectively \cite{Hadani2017}\cite{Monk2016OTFSO}.
A (coded) bit sequence is mapped to symbols $\{x[k,l], k = 0,\cdots,N-1, l =0,\cdots M-1\}$ in the DD domain, where $x[k,l]\in\mA=\{\alpha_1, ....\alpha_{|\mA|}\}$ with $|\mA|$ being the cardinality of $\mA$, and $l$ and $k$ denote the indices of the delay and Doppler shifts, respectively.  As shown in Fig. \ref{fig:model}, ISFFT is performed to convert the symbols to signals in the time-frequency (TF) domain, i.e.,
\begin{eqnarray}
X_{tf}[n,m] = \frac{1}{\sqrt{MN}}\sum_{k=0}^{N-1}\sum_{l=0}^{M-1}x[k,l]e^{j2\pi(\frac{nk}{N}-\frac{ml}{M})}.
\end{eqnarray}
Then $\{X_{tf}[m,n]\}$ are converted to a continuous-time waveform $s(t)$ using the Heisenberg transform with a transmit waveform $g_{tx}(t)$, i.e.,
\begin{align}
s(t) = \sum_{n=0}^{N-1}\sum_{m=0}^{M-1}X_{tf}[n, m]g_{tx}(t-nT)e^{j2\pi m\Delta f(t-nT)},
\end{align}
where $\Delta f$ is the subcarrier spacing and $T = 1/\Delta f$.

The signal $s(t)$ is then transmitted through a time-varying channel and the received signal in the time domain can be expressed as
\begin{align}
r(t) = \int\int h(\tau,\nu)s(t-\tau)e^{j2\pi\nu(t-\tau)} d\tau d\nu,
\end{align}
where $h(\tau, \nu)$ is the channel impulse response in the (continuous) DD domain. The channel impulse response can be expressed as
\begin{align}
h(\tau, \nu) = \sum_{i=1}^P h_i\delta(\tau-\tau_i)\delta(\nu - \nu_i),
\end{align}
where $\delta(\cdot)$ is the Dirac delta function, $P$ is the number of resolvable propagation paths, and $h_i$, $\tau_i$ and $\nu_i$ represent the gain, delay shift and Doppler shift associated with the $i$th path, respectively. The delay and Doppler shift taps for the $i$th path are
\begin{eqnarray}
\tau_i=\frac{l_i}{M\Delta f}, \\
\nu_i=\frac{k_i+\kappa_i}{NT}, \label{eq:ddtap}
\end{eqnarray}
where $0 \leq l_i \leq l_{max}$ and $-k_{max} \leq k_i \leq k_{max}$ are the delay index and Doppler index of the $i$th path, $l_{max}$ and $k_{max}$ represent the largest indices of the delay taps and Doppler taps, respectively, and $\kappa_{i}$ $\in[-0.5,0.5]$ is the fractional Doppler shift associated with the $i$th path.

At the receiver side, a receive waveform $g_{rx}(t)$ is used to transform the received signal $r(t)$ to the TF domain i.e.,
\begin{align}
Y(t,f) = \int g_{rx}^{*}(t'-t)r(t')e^{-j2\pi f(t'-t)}dt',
\end{align}
which is then sampled at $t=nT$ and $f=m\Delta f$, yielding $Y[n, m]$.
Finally, an SFFT is applied to $\{Y[n, m]\}$ to obtain the signal $y[k,l]$ in the DD domain, i.e.,
\begin{align}
y[k,l] = \frac{1}{\sqrt{MN}} \sum_{n=0}^{N-1}\sum_{m=0}^{M-1}Y[n,m]e^{-j2\pi(\frac{nk}{N} - \frac{ml}{M})}.
\end{align}

If the transmit waveform $g_{tx}(t)$ and receive waveform $g_{rx}(t)$ satisfy the bi-orthogonal property \cite{Hadani2017}, the channel input-output relationship in the DD domain can be expressed as \cite{Raviteja2018,Raviteja2019Practical}
\begin{eqnarray}
y[k,l]=\sum_{i=1}^P\sum_{q=-N_i}^{N_i}h_i\frac{1-e^{-j2\pi(-q-\kappa_i)}}{N-Ne^{-j\frac{2\pi}{N}(-q-\kappa_i)}}e^{-j2\pi \frac{l_i(k_i+\kappa_i)}{MN}} \nonumber \\
\times x\big[[k-k_i+q]_N,[l-l_i]_M\big]
  + \omega[k,l],\label{eq:IdealFracY}
\end{eqnarray}
where
$N_i\ll N$ is an integer, and $ \omega[k,l]$ denotes the Gaussian noise in the DD domain with mean 0 and variance $\gamma^{-1}$ (or precision $\gamma$). We can see that for each path, the transmitted signal is circularly shifted, and scaled by a channel gain. We stack $\{x[k,l]\}$ to form a vector $\boldsymbol{x}\in \mathbb{C}^{MN \times 1}$, where the $j$th element $x_j$ is $x[k,l]$ with $j = kM + l$. Similarly, a vector $\boldsymbol{y}\in \mathbb{C}^{MN \times 1}$ can also be constructed from $\{y[k,l]\}$. Then \eqref{eq:IdealFracY} can be rewritten in \rev{a} vector form as
\begin{eqnarray}
\boldsymbol{y} = \boldsymbol{H}_{bi}\boldsymbol{x} + \boldsymbol{\omega}, \label{eq:idealVectorFracY}
\end{eqnarray}
where $\boldsymbol{\omega}$ is the corresponding noise vector, and $\boldsymbol{H}_{bi} \in \mathbb{C}^{MN \times MN}$ represents the effective channel in the DD domain, which can be expressed as
\begin{eqnarray}
\boldsymbol{H}_{bi} = \sum_{i=1}^{P}\sum_{q=-N_i}^{N_i}\boldsymbol{I}_N(-[q-k_i]_N)\otimes \Big[\boldsymbol{I}_M(l_i)h_i \nonumber \\ \times  \left(\frac{1 - e^{-j2\pi(-q-\kappa_i)}}{N - Ne^{-j\frac{2\pi}{N}(-q-\kappa_i)}}\right)e^{-j2\pi\frac{l_i(k_i+\kappa_i)}{MN}}\Big], \label{eq:idealDDH}
\end{eqnarray}
where $\boldsymbol{I}_N(-[q-k_i]_N)$ denotes an $N\times N$ matrix obtained by circularly shifting the rows of an identity matrix by $-[q-k_i]_N$, and $\boldsymbol{I}_M(l_i)$ is obtained similarly.

When the rectangular waveform is used for both $g_{tx}(t)$ and $g_{rx}(t)$, the received signal $y[k,l]$ in the DD domain can be expressed as \cite{Raviteja2018,Raviteja2019Practical}
\begin{eqnarray}
y[k,l]=\sum_{i=1}^P\sum_{q=-N_i}^{N_i}h_i e^{j2\pi\left(\frac{l-l_i}{M}\right)\left(\frac{k_i +\kappa_i}{N}\right)}\alpha_i(k,l,q)\nonumber \\
 \times x[[k-k_i+q]_N, [l-l_i]_M] + \omega[k,l], \label{eq:RectFracY}
\end{eqnarray}
where
\begin{equation}
    \alpha_i(k,l,q)=\begin{cases}
    \frac{1}{N}\beta_i(q) & l_i\leq l < M \\
    \frac{1}{N}(\beta_i(q)-1) e^{-j2\pi\frac{[k-k_i+q]_N}{N}} & 0\leq l < l_i
    \end{cases}, \label{eq:rectalpha}
\end{equation}
and
\begin{eqnarray}
\beta_i(q)=\frac{e^{-j2\pi(-q-\kappa_i)}-1}{e^{-j\frac{2\pi}{N}(-q-\kappa_i)}-1}.
\end{eqnarray}

Similarly, we can also rewrite (\ref{eq:RectFracY}) in a vector form as
\begin{align}
\boldsymbol{y} = \boldsymbol{H}_{rect}\boldsymbol{x}+\boldsymbol{\omega},
\end{align}
and the channel matrix $\boldsymbol{H}_{rect} \in \mathbb{C}^{MN \times MN}$ can be written as
\begin{align}
  \boldsymbol{H}_{rect}=& \sum_{i=1}^{P}\sum_{q=-N_i}^{N_i}\boldsymbol{I}_N(-[q-k_i]_N)\otimes \nonumber \\ &\left(\boldsymbol{\Lambda}\boldsymbol{I}_M(l_i)h_i  e^{-j2\pi\frac{l_i(k_i+\kappa_i)}{MN}}\right) \cdot\boldsymbol{\Delta}^{-[q-k_i]_N},
\end{align}
where $\boldsymbol{\Lambda}\in \mathbb{C}^{M\times M}$ is a diagonal matrix and the $l$th diagonal element $\boldsymbol{\Lambda}_{ll}$ can be expressed as
\begin{figure}[htbp]
	\centering
	\includegraphics[width=0.95\columnwidth]{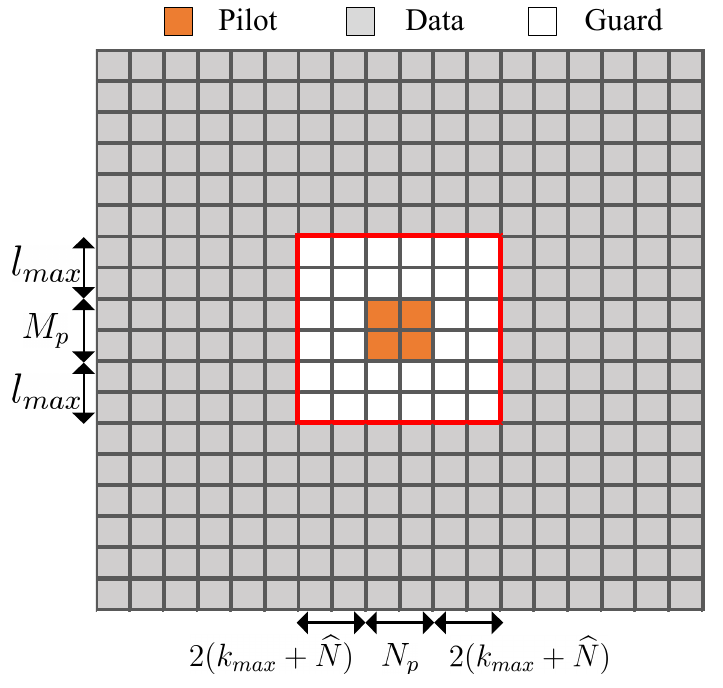}
	\caption{An OTFS block with pilot symbols inserted in the DD domain \cite{chestmassivemimo}.}
	\label{fig:txpattern}
\end{figure}

\begin{align}
\boldsymbol{\Lambda}_{ll}=
\begin{cases}
e^{j2\pi\frac{l(k_i+\kappa_i)}{MN}}\beta_i(q)/N & l_i\leq l < M \\
e^{j2\pi\frac{l(k_i+\kappa_i)}{MN}}\left(\beta_i(q) - 1\right)/N  & 0\leq l < l_i\\
\end{cases}.
\end{align}
Note that $\boldsymbol{\Delta}^{-[q-k_i]_N}$ is an $MN\times MN$ block matrix, which is obtained by circularly shifting the blocks in a matrix $\boldsymbol{\Delta}$ by $-[q-k_i]_N$. The matrix $\boldsymbol{\Delta}$  is a block diagonal matrix
\begin{align}
  \boldsymbol{\Delta} = \left (\begin{array}{cccccc}
\boldsymbol{\Delta}_0 &\boldsymbol{0}   & \cdots& \boldsymbol{0} \\
\boldsymbol{0} &\boldsymbol{\Delta}_1 & \cdots & \boldsymbol{0} \\
\vdots & \vdots &\ddots & \vdots\\
\boldsymbol{0} & \boldsymbol{0} &\cdots & \boldsymbol{\Delta}_{N-1}\\
\end{array}\right),
\end{align}
where $\boldsymbol{\Delta}_n = \boldsymbol{\Psi I}_M(l_i)$ and $\boldsymbol{\Psi}\in \mathbb{C}^{M\times M}$ is a diagonal matrix with the $m$th diagonal element ${\Psi}_{mm}$  given as
\begin{align}
{\Psi}_{mm}=
\begin{cases}
1 & l_i\leq m < M \\
e^{-j2\pi m/N}  & 0\leq m < l_i\\
\end{cases}.
\end{align}

\begin{figure}[htbp]
	\centering
	\includegraphics[width=0.9\columnwidth]{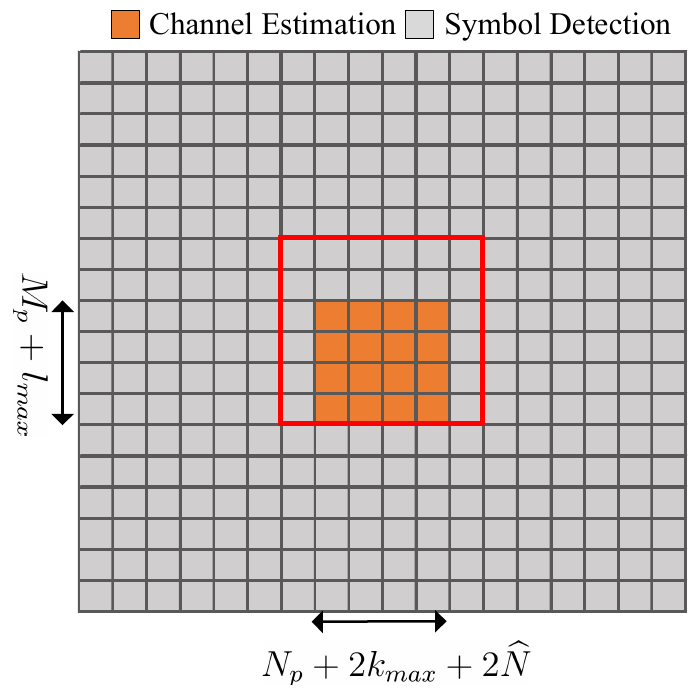}
	\caption{Received OTFS signal block where the signals in the orange area are used for channel estimation \cite{chestmassivemimo}.}
	\label{fig:rxpattern}
\end{figure}

\subsection{Data and Pilot Symbols Arrangement}
In this paper, we use the scheme in \cite{pilotref} and \cite{chestmassivemimo}, where pilot symbols are inserted to the DD plane with guard interval to avoid the interference between data and pilot symbols, \rev{as} shown in Fig. \ref{fig:txpattern}. This allows that channel estimation and data detection are carried out for the same OTFS block.

As shown in Fig. \ref{fig:txpattern}, the size of the pilot symbol block is $M_p \times N_p$. To avoid the interference between pilot and data symbols, the size of guard interval should be at least $2(k_{max} + \widehat{N})$ along the Doppler dimension and $l_{max}$ along the delay dimension, where we assume $\{N_i, i=1,2,\cdots, P\}$ take a same value $\widehat{N}$. \rev{With this scheme, the overhead due to pilot symbols and guard interval is $(2l_{max}+M_p)(4k_{max}+4\widehat{N}+N_p)/MN$.}  
The received signal in the DD domain is shown in Fig. \ref{fig:rxpattern}, where the signals in the orange area correspond to the pilot symbols, and \rev{they} are used for channel estimation.

{In order to minimize the spectrum overhead due to pilot symbols, the use of a small number of pilot symbols is highly desirable. To guarantee the channel estimation quality, the power of the pilot symbols should be sufficiently high, which can be much larger than the power of data symbols. Fortunately, the operation of ISFFT at the transmitter can spread the \rev{power of} pilot symbols, therefore \rev{greatly} alleviating the PAPR problem \cite{pilotref}. In \cite{pilotref}, 	
a single pilot symbol with a very high power is used \rev{to facilitate the design of a low complexity channel estimation algorithm based on thresholding. However, the use of a few pilot symbols can be more preferable because this can significantly reduce the PAPR at the cost of a small loss in spectrum efficiency.} In Table \ref{tab:papr}, we compare the average PAPR of OTFS signals in the time domain, where we assume that the rectangle waveform is used. The pilot symbols are arranged in a single column and placed along the delay dimension, i.e., $N_p=1$. We assume that the signal to noise ratio (SNR) of data symbols is 14 dB, and compare the average PAPR of the OTFS signals versus the number of pilot symbols and the SNR of pilot symbols (denoted by SNRp). It can be seen that more pilot symbols are used, a lower PAPR can be achieved. \rev{Moreover, with a fixed power budget for pilot,} the use of multiple pilot symbols leads to considerable PAPR reduction. \rev{Take an example: the use of a single pilot symbol with SNRp = 50dB has the same power budget as the use of 10 pilot symbols with SNRp = 40dB. From Table \ref{tab:papr}, a significant PAPR reduction of about 8dB can be achieved. It is worth mentioning that, with the proposed method in this paper, almost the same estimation performance can be achieved for both cases (as shown later), making the use of multiple pilot symbols very attractive.}

\begin{table}[htb]
\centering
\renewcommand\arraystretch{1.4}
\caption{PAPR of OTFS signals in the time domain, where $M=128$, $N=32$ and the SNR of data symbol is 14 dB.}\label{tab:papr}
\begin{tabular}{|>{\centering}p{40pt}|>{\centering}p{100pt}|>{\centering \arraybackslash }p{70pt}|}
\hline
SNRp & \# of pilot symbols & PAPR  \\
\hline
\multirow{2}*{40dB}  & 1 & 12.5527 dB\\
\cline{2-3}
  ~  & 10 & 10.4095 dB\\
\hline
  \multirow{2}*{50dB}  & 1 & 18.7619 dB\\
\cline{2-3}
  ~  & 10 & 11.3900 dB\\
\hline
\end{tabular}
\end{table}

In this work, we design a flexible channel estimation method, which is able to work with a single or multiple pilot symbols. Moreover, the proposed method is able to estimate the channel gains and the fractional Doppler shifts, which are crucial to improving the estimation performance as demonstrated later. 

\section{Structured Sparse Signal Recovery For OTFS Channel Estimation} \label{sec:formulation}

The OTFS channel can be estimated using the pilot symbols and the received signals in the orange area shown in Fig. \ref{fig:rxpattern}. It can be seen from (\ref{eq:IdealFracY}) and (\ref{eq:RectFracYSimplified}) that the OTFS channel is parameterized by the nonzero channel gains $\{h_i\}$, the fractional Doppler shifts $\{\kappa_i\}$ and their indices $\{l_i, k_i\}$, $i=1,\cdots,P$. These parameters will be estimated, based on which the OTFS channel in the DD domain can be constructed.

\subsection{Bi-Orthogonal Waveform} \label{section:BiWave}
{Model \eqref{eq:IdealFracY} involves a number of unknown variables, including $h_i$, $\kappa_i$, $l_i$ and $k_i$ for $i=1,\cdots,P$. In addition, $P$ is unknown as well. These make the estimation very difficult. To overcome this, we define two variables $t$ and $d$ to represent the indices of delay shifts and Doppler shifts, respectively. As only the pilot symbols need to be considered, we have $t \in [0, l_{max} ]$ and $d \in [-k_{max}, k_{max} ]$. Then \eqref{eq:IdealFracY} can be rewritten as
\begin{eqnarray}
y[k,l]=\sum_{t=0}^{l_{max}} \sum_{d=-k_{max}}^{k_{max}} h_{t,d}e^{-j2\pi \frac{t(d+\kappa_{d})}{MN}}\sum_{q=-\widehat{N}}^{\widehat{N}}f(q,\kappa_{d}) \nonumber \\
 x\big[[k-d+q]_N,[l-t]_M\big]
  + \omega[k,l],  \label{eq:IdealRecvFracY}
\end{eqnarray}
where
\begin{eqnarray}
f(q,\kappa_{d})=\frac{1-e^{-j2\pi(-q-\kappa_{d})}}{N-Ne^{-j\frac{2\pi}{N}(-q-\kappa_{d})}}. \label{eq:fqkappa}
\end{eqnarray}
The received signals $\{y[k,l]\}$  in (\ref{eq:IdealRecvFracY}) corresponding to the pilot symbols are in the orange area in Fig. \ref{fig:rxpattern}.
Our aim is to find the non-zero elements in $\{h_{t,d}, \kappa_{d}, t \in [0, l_{max} ], d \in [-k_{max}, k_{max}] \}$ based on $\{y[k,l]\}$. Therefore, the estimation can be formulated as a sparse signal recovery problem.

To facilitate the design of the estimation algorithm, we rewrite \eqref{eq:IdealRecvFracY} in a matrix form as
\begin{eqnarray}
\boldsymbol{y} = \rev{\boldsymbol{X}_{bi}}\boldsymbol{c}+\boldsymbol{\omega}, \label{eq:idealVectorY}
\end{eqnarray}
where $\boldsymbol{y} \in \mathbb{C}^{Z\times 1}$, $Z=(l_{max}+M_p)(N_p + 2k_{max}+2\widehat{N})$ is formed by stacking $\{y[k,l]\}$ as a vector, $\rev{\boldsymbol{X}_{bi}} \in \mathbb{C}^{Z\times (l_{max}+1)(2k_{max}+1)(2\widehat{N}+1)}$ is constructed based on the pilot symbols, $\boldsymbol{\omega} \in \mathbb{C}^{Z\times 1}$ is obtained by stacking $\{\omega[k,l]\}$, and $\boldsymbol{c}\in \mathbb{C}^{(l_{max}+1)(2k_{max}+1)B\times 1}$ with $B=2\widehat{N}+1$ can be expressed as
\begin{align}
\boldsymbol{c} = [\boldsymbol{c}^T_{0,-k_{max}}, ~\boldsymbol{c}^T_{1,-k_{max}}, ~\cdots, ~\boldsymbol{c}^T_{t, d},   \cdots,  \boldsymbol{c}^T_{l_{max},k_{max}}]^T, \label{eq:cvecdef}
\end{align}
where $\boldsymbol{c}_{t, d} \in \mathbb{C}^{B \times 1}$ is given as
\begin{eqnarray}
&&\boldsymbol{c}_{t,d} = h_{t,d}e^{-j2\pi\frac{(d+\kappa_d)t}{MN}}\cdot \nonumber \\
&& \ \ [f(-\widehat{N}, \kappa_d),f(-\widehat{N}+1, \kappa_d),\cdots, f(\widehat{N}, \kappa_d)]^T.
\end{eqnarray}

\rev{We can see that the vector $\boldsymbol{c}$ has a structure}, i.e., $\boldsymbol{c}$ is block sparse, and for a non-zero block (subvector) $\boldsymbol{c}_{t,d}$, it is parameterized by only two parameters $h_{t,d}$ and $\kappa_d$.
We will recover the structured sparse vector $\boldsymbol{c}$ and obtain the estimates of the non-zero channel gains and fractional Doppler shifts.

%

\subsection{Rectangular Waveform}

To facilitate channel estimation, we place the pilot symbols to ensure that their delay index $l \geq l_{max}$, so that (\ref{eq:RectFracY}) is reduced to
\begin{eqnarray}
y[k,l]=\!\!\!\!\!\!\!\!&&\sum_{i=1}^P\sum_{q=-N_i}^{N_i}h_i\frac{1-e^{-j2\pi(-q-\kappa_i)}}{N-Ne^{-j\frac{2\pi}{N}(-q-\kappa_i)}}e^{j2\pi \frac{(l-l_i)(k_i+\kappa_i)}{MN}} \nonumber \\
&& ~~\times x\big[[k-k_i+q]_N,[l-l_i]_M\big] + \omega[k,l]. \label{eq:RectFracYSimplified}
\end{eqnarray}
Comparing (\ref{eq:RectFracYSimplified}) to (\ref{eq:IdealFracY}), we can find that their difference lies in that \rev{each transmitted symbol 
in (\ref{eq:RectFracYSimplified}) }is rotated by an additional phase $e^{j2\pi \frac{l(k_i+\kappa_i)}{MN}}$.


Similar to the case of bi-orthogonal waveform, model (\ref{eq:RectFracYSimplified}) can be rewritten in a matrix form
\begin{align}
\boldsymbol{y} = \rev{\boldsymbol{X}_{rect}}\boldsymbol{c}+\boldsymbol{\omega}, \label{eq:rectVectorY}
\end{align}
where $\boldsymbol{c}$ is the structured sparse vector to be recovered, and the $(z,n)$th element of matrix $\boldsymbol{X}_{rect}$ is given as
\begin{align}
  \rev{{X}_{rect}^{z,n} = e^{j2\pi\frac{l_z(d_n+\kappa_n)}{MN}}{X}_{bi}^{z,n}} \label{eq:rectX},
\end{align}
where $l_z$ is the delay index corresponding to the $z$th element of $\boldsymbol{y}$, $d_n$ and $\kappa_n$ are the parameters $d$ and $\kappa_d$ of the $n$th element of vector $\boldsymbol{c}$ (refer to the definition of vector $\boldsymbol{c}$ in \eqref{eq:cvecdef}, \rev{and ${X}_{bi}^{z,n}$ denotes the the $(z,n)$th element of matrix $\boldsymbol{X}_{bi}$}. It is noted that when $z$ and $n$ are given, 
 $l_z$ and $d_n$ are known, but $\kappa_n$ is unknown. So it is different from the case of the bi-orthogonal waveform in (\ref{eq:idealVectorY}) that the matrix \rev{$\boldsymbol{X}_{rect}$} depends on the fractional Doppler shifts, which are unknown. Later we will show that this can be solved by using an iterative estimation strategy.


As discussed above, the channel estimation problem can be formulated as structured sparse signal recovery. Next, we will develop a Bayesian method to recover the block sparse vector $\boldsymbol{c}$ and obtain the estimates of $\{h_{t,d}\}$ and $\{\kappa_d\}$.
In particular, the factor graph techniques \cite{Kschischang2001} are used, based on which an efficient message passing algorithm is developed.

\section{Bayesian Approach and Message Passing Algorithm} \label{sec:algorithm}

In this section, we first focus on OTFS with the bi-orthogonal waveform, and then extend our discussion to the case of the rectangular waveform.

For the convenience of notation, we define $j = (l_{max}+1)(k_{max}+d)+t+1$. As $t \in [0, l_{max}]$ and $d \in [-k_{max}, k_{max}]$, there is one-to-one map between the index $j$ and the index pair $(t,d)$. Then we define
\begin{align}
    \boldsymbol{c}_j = \boldsymbol{c}_{t,d}, ~ h_j = h_{t,d}, ~ \kappa_j = \kappa_d,
\end{align}
which will be used hereafter.

Inspired by the sparse Bayesian learning \cite{Tipping2001Sparse}, we assume a Gaussian prior with mean 0 and precision $\lambda_j$ for $h_j$ to promote sparsity, i.e.,
\begin{align}
    p(h_j|\lambda_j) = \mathcal{CN}(h_j; 0, \lambda_j^{-1}),
\end{align}
where the hyperparameter $\lambda_j$ is Gamma distributed, i.e., $p(\lambda_j)=Ga(\lambda_j;\epsilon_j, \eta_j)$. \rev{The hyper-parameters $\epsilon_j$ and $\eta_j$ are respectively set to be 1 and 0, leading to a uniform or noninformative distribution for $\lambda_j$ \cite{cslaplace}.} As the noise precision $\gamma$ is normally unknown, it will also be estimated with an improper prior $p(\gamma)$\rev{\cite{Tipping2001Sparse}}. 

To facilitate the algorithm design, {we introduce an auxiliary vector
	\rev{
\begin{align}
\boldsymbol{g}_j \triangleq&  [f(-\widehat{N}, \kappa_j),f(-\widehat{N}+1, \kappa_j),\cdots, f(\widehat{N}, \kappa_j)]^T e^{-j2\pi\frac{t(d+\kappa_j)}{MN}} \nonumber \\
\triangleq & [\Phi(-\widehat{N}, \kappa_j), \cdots, \Phi(q, \kappa_j), \cdots, \Phi(\widehat{N}, \kappa_j)]^T, \label{eq:nuj}
\end{align}
}
where
\begin{equation}
\Phi(q, \kappa_j)=f(q,\kappa_j)e^{-j2\pi\frac{t(d+\kappa_j)}{MN}}.
\end{equation}
Then we have
\begin{align}
\boldsymbol{c}_j = h_j \boldsymbol{g}_j.
\end{align}
Define $\boldsymbol{h} =[h_1,\cdots h_J]^T$, $\boldsymbol{g}=[\boldsymbol{g}_1^T, \cdots, \boldsymbol{g}_J^T ]^T$,  $\boldsymbol{\kappa}=[\kappa_1, \cdots, \kappa_J]^T$ and  $\boldsymbol{\lambda}=[\lambda_1, \cdots, \lambda_J]^T$, where $J=(l_{max}+1)(2k_{max}+1)$. Then the joint conditional distribution of the unknown variables can be factorized as
\begin{align}
 p&(\boldsymbol{c}, \boldsymbol{h}, \boldsymbol{g}, \boldsymbol{\kappa},\boldsymbol{\lambda}, \gamma |\boldsymbol{y}) \nonumber \\ \propto &  ~p(\boldsymbol{y}|\boldsymbol{c},\gamma)p(\boldsymbol{c}|\boldsymbol{h},\boldsymbol{g})p(\boldsymbol{h}|\boldsymbol{\lambda})p(\boldsymbol{\lambda})p(\boldsymbol{g}|\boldsymbol{\kappa})p(\boldsymbol{\kappa})p(\gamma) \nonumber\\
 =&~ p(\gamma) p(\boldsymbol{y}|\boldsymbol{c},\gamma){\prod}_{j,b}p(c_{jb}|{h}_j,g_{jb})p(h_j|\lambda_j)p(\lambda_j)\nonumber \\ &\qquad\qquad\times p(g_{jb}|\kappa_j)p(\kappa_j)\nonumber \\
 \triangleq &~ f_{\gamma}(\gamma)f_{\boldsymbol{y}}(\boldsymbol{c},\gamma){\prod}_{j,b}f_{c_{jb}}(c_{jb},h_j,g_{jb})f_{h_j}(h_j, \lambda_j)f_{\lambda_j}(\lambda_j) \nonumber \\ & \qquad\qquad \times f_{g_{jb}}(g_{jb},\kappa_j)f_{\kappa_j}(\kappa_j),\label{eq:idealFactor}
\end{align}
%
\rev{where $1\leq j \leq J$, $1 \leq b \leq B$, and $c_{jb}$ and $g_{jb}$ represent the $b$th element of $\boldsymbol{c}_j$ and $\boldsymbol{g}_j$, respectively. The correspondence between the factors and distributions (and their detailed forms) are listed in Table \ref{tab:factor}.} \rev{We aim to obtain the approximate marginals of the parameters (thereby their estimates) by performing approximate inference. In particular, we use the factor graph techniques, and the variational message passing (VMP) \cite{VAMP, MF2002} and belief propagation (BP) message passing \cite{Kschischang2001} are combined to achieve efficient approximate inference. We note that variational inference has been used for channel estimation, e.g., the works in \cite{Reviewer3R1} and \cite{Reviewer3R2}.}  

\tikzstyle{factornode} = [draw, fill=white, circle, inner sep=1pt,minimum size=0.8cm]
\tikzstyle{funnode} = [draw, rectangle,fill=black!100, minimum size = 0.6cm]
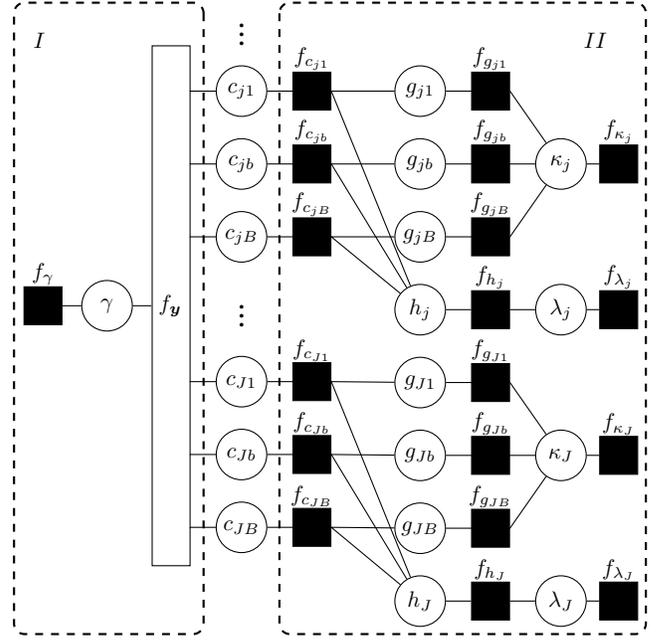
\begin{figure}[htbp]
	\centering
\begin{tikzpicture} [scale=0.84, transform shape]	
	\node (cj1)[factornode] at (0,0) {$c_{j1}$};
	\node (cjb)[factornode, below = 0.34cm of cj1]  {$c_{jb}$};
	\node (cjB)[factornode, below = 0.34cm of cjb]  {$c_{jB}$};

	\node (vdot1)[above = 0.2cm of cj1]  {$\boldsymbol{\vdots}$};
	\node (vdot2)[below = 0.34cm of cjB]  {$\boldsymbol{\vdots}$};
	
	\node (cJ1)[factornode, below = 0.34cm of vdot2]  {$c_{J1}$};
	\node (cJb)[factornode, below = 0.34cm of cJ1]  {$c_{Jb}$};
	\node (cJB)[factornode, below = 0.34cm of cJb]  {$c_{JB}$};
		
    \node (fcj1)[right = 0.4cm of cj1, funnode]  {};
    \node (fcjb)[right = 0.4cm of cjb, funnode]  {};
	\node (fcjB)[right = 0.4cm of cjB, funnode]  {};
	\node (fcJ1)[right = 0.4cm of cJ1, funnode]  {};
	\node (fcJb)[right = 0.4cm of cJb, funnode]  {};
	\node (fcJB)[right = 0.4cm of cJB, funnode]  {};
	
	\node [above= -0.1cm of fcj1]{$f_{c_{j1}}$};
	\node [above= -0.1cm of fcjb]{$f_{c_{jb}}$};
	\node [above= -0.1cm of fcjB]{$f_{c_{jB}}$};
	\node [above= -0.1cm of fcJ1]{$f_{c_{J1}}$};
	\node [above= -0.1cm of fcJb]{$f_{c_{Jb}}$};
	\node [above= -0.1cm of fcJB]{$f_{c_{JB}}$};

	\node (gj1)[factornode,  right = 1cm of fcj1]  {$g_{j1}$};
	\node (gjb)[factornode, below = 0.34cm of gj1]  {$g_{jb}$};
	\node (gjB)[factornode, below = 0.34cm of gjb]  {$g_{jB}$};
	\node (hj)[factornode, below = 0.34cm of gjB]  {$h_j$};
	
	\node (gJ1)[factornode, below = 0.34cm of hj]  {$g_{J1}$};
	\node (gJb)[factornode, below = 0.34cm of gJ1]  {$g_{Jb}$};
	\node (gJB)[factornode, below = 0.34cm of gJb]  {$g_{JB}$};
	\node (hJ)[factornode, below = 0.34cm of gJB]  {$h_J$};

	\node (fgj1)[funnode, right = 0.4cm of gj1]  {};
	\node (fgjb)[funnode, right = 0.4cm of gjb]  {};
	\node (fgjB)[funnode, right = 0.4cm of gjB]  {};
	\node (fhj)[funnode, right = 0.4cm of hj]  {};
	
	\node (fgJ1)[funnode, right = 0.4cm of gJ1]  {};
	\node (fgJb)[funnode, right = 0.4cm of gJb]  {};
	\node (fgJB)[funnode, right = 0.4cm of gJB]  {};
	\node (fhJ)[funnode, right = 0.4cm of hJ]  {};
	
	\node [above= -0.1cm of fgj1]{$f_{g_{j1}}$};
	\node [above= -0.1cm of fgjb]{$f_{g_{jb}}$};
	\node [above= -0.1cm of fgjB]{$f_{g_{jB}}$};
	\node [above= -0.1cm of fhj]{$f_{h_j}$};

	\node [above= -0.1cm of fgJ1]{$f_{g_{J1}}$};
	\node [above= -0.1cm of fgJb]{$f_{g_{Jb}}$};
	\node [above= -0.1cm of fgJB]{$f_{g_{JB}}$};
	\node [above= -0.1cm of fhJ]{$f_{h_J}$};
	
	\node (kappaj)[factornode, right = 0.4cm of fgjb]  {$\kappa_j$};
	\node (lambdaj)[factornode, right = 0.4cm of fhj]  {$\lambda_j$};
	\node (kappaJ)[factornode, right = 0.4cm of fgJb]  {$\kappa_J$};
	\node (lambdaJ)[factornode, right = 0.4cm of fhJ]  {$\lambda_J$};
	
	\node (fkappaj)[funnode, right = 0.2cm of kappaj]  {};
	\node (flambdaj)[funnode, right = 0.2cm of lambdaj]  {};
	\node (fkappaJ)[funnode, right = 0.2cm of kappaJ]  {};
	\node (flambdaJ)[funnode, right = 0.2cm of lambdaJ]  {};
	
	\node [above= -0.1cm of fkappaj]{$f_{\kappa_j}$};
	\node [above= -0.1cm of flambdaj]{$f_{\lambda_j}$};
	\node [above= -0.1cm of fkappaJ]{$f_{\kappa_J}$};
	\node [above= -0.1cm of flambdaJ]{$f_{\lambda_J}$};
	
	\node (fy)[rectangle,draw, left = 0.4cm of cj1, text height=8cm, minimum size = 0.6cm, yshift=-3.4cm, align=center] {};
	\node (fytext)[left = -0.6cm of fy, align=center] {$f_{\boldsymbol{y}}$};
	\node (gamma)[factornode, left = 0.3cm of fy]  {$\gamma$};
	\node (fgamma)[funnode, left = 0.3cm of gamma]  {};
	\node [above= -0.1cm of fgamma]{$f_{\gamma}$};
	
	\node (part1) at (5.6, 0.8) {$\uppercase\expandafter{\romannumeral2}$};
	\node (part1) at (-3.2, 0.8) {$\uppercase\expandafter{\romannumeral1}$};
	
	\draw (cj1) -- (fcj1);
	\draw (cjb) -- (fcjb);
	\draw (cjB) -- (fcjB);
	\draw (cJ1) -- (fcJ1);
	\draw (cJb) -- (fcJb);
	\draw (cJB) -- (fcJB);
	
	\draw (gj1) -- (fcj1.east) -- (hj);
	\draw (gjb) -- (fcjb.east) -- (hj);
	\draw (gjB) -- (fcjB.east) -- (hj);
	
	\draw (gJ1) -- (fcJ1.east) -- (hJ);
	\draw (gJb) -- (fcJb.east) -- (hJ);
	\draw (gJB) -- (fcJB.east) -- (hJ);
	
	\draw (gj1) -- (fgj1.west) -- (fgj1.east) -- (kappaj);
	\draw (gjb) -- (fgjb.west) -- (fgjb.east) -- (kappaj);
	\draw (gjB) -- (fgjB.west) -- (fgjB.east) -- (kappaj);
	\draw (hj) -- (fhj) -- (lambdaj) -- (flambdaj);
	
	\draw (gJ1) -- (fgJ1.west)-- (fgJ1.east) -- (kappaJ);
	\draw (gJb) -- (fgJb.west)-- (fgJb.east) -- (kappaJ);
	\draw (gJB) -- (fgJB.west)-- (fgJB.east) -- (kappaJ);
	\draw (hJ) -- (fhJ) -- (lambdaJ) -- (flambdaJ);
	
	\draw (kappaj) -- (fkappaj);
	\draw (kappaJ) -- (fkappaJ);
	
	\draw (fgamma) -- (gamma);
	\draw (gamma) -- ($(gamma) + (0.7,0)$);
	\draw (cj1) -- ($(cj1) + (-0.82,0)$);
	\draw (cjb) -- ($(cjb) + (-0.82,0)$);
	\draw (cjB) -- ($(cjB) + (-0.82,0)$);
	\draw (cJ1) -- ($(cJ1) + (-0.82,0)$);
	\draw (cJb) -- ($(cJb) + (-0.82,0)$);
	\draw (cJB) -- ($(cJB) + (-0.82,0)$);

	\draw[thick, dashed, rounded corners] ($(0.6,1.4)$) rectangle ($(6.4,-8.6)$);
	\draw[thick, dashed, rounded corners] ($(-0.6,1.4)$) rectangle ($(-3.6,-8.6)$);	
\end{tikzpicture}
	\caption{\rev{ Factor graph representation of \eqref{eq:idealFactor}.}}
\label{fig:idealFactorGraph}
\end{figure}

\begin{table}[htb]
	\color{black} 
	\centering
	\renewcommand\arraystretch{1.2}
	\caption{\rev{Correspondence between the factors and distributions in (\ref{eq:idealFactor})}.}\label{tab:factor}
	\begin{tabular}{>{\centering}p{30pt}>{\centering}p{60pt} >{\centering \arraybackslash }p{120pt}}
		\hline
		Factor & Distribution & Function  \\
		\hline
		$f_{\gamma}$  & $p(\gamma)$ & $\gamma^{-1}$ \\	
		
		$f_{\boldsymbol{y}}$  & $p(\boldsymbol{y}|\boldsymbol{c},\gamma)$ & $\mathcal{CN}(\boldsymbol{y}; \boldsymbol{X}_{bi}\boldsymbol{c},\gamma^{-1}\boldsymbol{I}_Z)$ \\
		
		$f_{c_{jb}}$ & $p(c_{jb}|h_j, g_{jb})$ & $\delta(c_{jb} - h_jg_{jb})$ \\
		$f_{h_j}$ & $p(h_j|\lambda_j)$ & $\mathcal{CN}(h_j; 0, \lambda_j^{-1})$ \\
		$f_{\lambda_j}$ & $p(\lambda_j)$ & $Ga(\lambda_j;\epsilon_j, \eta_j)$ \\
		$f_{g_{jb}}$ & $p(g_{jb}|\kappa_j)$ & $\delta(g_{jb} - \Phi(-\widehat{N}+b-1,\kappa_j))$ \\
		$f_{\kappa_j}$ & $p(\kappa_j)$ & $U[-0.5,0.5]$ \\
		\hline
	\end{tabular}
\end{table}

The factorization of (\ref{eq:idealFactor}) can be visualized by the factor graph shown in Fig. \ref{fig:idealFactorGraph}, where we partition the factor graph into two parts: Part \uppercase\expandafter{\romannumeral1} and Part \uppercase\expandafter{\romannumeral2}. 
\rev{To keep the graph clear, we only show the function nodes and variables nodes associated with the $j$th and $J$th blocks of $\boldsymbol{c}$}.
In the following, we derive the message computations for the forward (from left to right) and backward (from right to left) passing in both parts. We use $m_{A\rightarrow B}(x)$ to denote a message passed from a function node $A$ to a variable node $B$, which is a function of $x$, and use $n_{B\rightarrow A}(x)$ to denote a message passed from a variable node $B$ to a function node $A$ \rev{, which is also a function of $x$}. Meanwhile, the arrows above the mean and the variance of a Gaussian message indicate the direction of the message passing. In addition, we use $b(x)$ to denote the belief of a variable $x$. Note that, if a forward computation requires backward messages, the relevant messages in the previous iteration is used by default.

\subsection{Message Computations in Part \uppercase\expandafter{\romannumeral1}}
Assume that the belief of $\boldsymbol{c}$ is known, which turns out to be Gaussian, i.e., $b(\boldsymbol{c})\propto \mathcal{CN}(\boldsymbol{c};\boldsymbol{c}^p,\boldsymbol{V}_c^p)$, as given later in (\ref{eq:beliefcvar}) and (\ref{eq:beliefcmean}). {The VMP or mean field (MF)}
is used at the function node $f_{\boldsymbol{y}}(\boldsymbol{c},\gamma)$, and the message from $f_{\boldsymbol{y}}(\boldsymbol{c},\gamma)$ to the variable node $\gamma$ can be computed as 
\begin{align}
m_{f_{\boldsymbol{y}}\rightarrow \gamma} (\gamma) &\propto \exp \int \ln f_{\boldsymbol{y}}(\boldsymbol{c},\gamma)b(\boldsymbol{c})d\boldsymbol{c} \nonumber\\
&\propto  {\gamma^Z}\exp\{-\gamma[(\boldsymbol{y}-\boldsymbol{X}_{bi}\boldsymbol{c}^p)^H(\boldsymbol{y}-\boldsymbol{X}_{bi}\boldsymbol{c}^p) \nonumber \\ & \qquad \qquad + Tr\{\boldsymbol{X}_{bi}\boldsymbol{V_c}^p\boldsymbol{X}_{bi}^H\}]\}.
\end{align}
With the prior $f_{\gamma}(\gamma) \propto 1/\gamma$, the belief $b(\gamma)$ can be expressed as
\begin{align}
 b(\gamma) \propto \gamma^{-1}m_{f_{\boldsymbol{y}}\rightarrow \gamma }(\gamma).
\end{align}
Then, with the MF rule and the belief $b(\gamma)$, we can compute the outgoing message $m_{f_{\boldsymbol{y}}\rightarrow {c}_{jb}}(c_{jb})$ as
\begin{align}
&m_{f_{\boldsymbol{y}}\rightarrow{c}_{jb}}(c_{jb}) \nonumber \\ &\propto \exp\int \ln f_{\boldsymbol{y}}(\boldsymbol{c},\gamma)b(\gamma)b(\boldsymbol{c}_{\sim c_{jb}})d\gamma d\boldsymbol{c}_{\sim c_{jb}},
\end{align}
where $\boldsymbol{c}_{\sim c_{jb}}$ denotes a vector obtained by removing $c_{jb}$ from $\boldsymbol{c}$.
However, the computation of the message involves high complexity. An efficient way is to first compute the belief of $c_{jb}$, then compute the outgoing extrinsic message. According to the MF rule, the message $m_{f_{\boldsymbol{y}}\rightarrow\boldsymbol{c}}(\boldsymbol{c})$ can be expressed as
\begin{align}
    &m_{f_{\boldsymbol{y}}\rightarrow\boldsymbol{c}}(\boldsymbol{c})\nonumber \\ &\propto \exp \int \ln f_{\boldsymbol{y}}(\boldsymbol{c},\gamma)b(\gamma)d\gamma \nonumber\\
    &\propto ~\mathcal{CN}\left(\boldsymbol{c}; (\boldsymbol{X}_{bi}^H\boldsymbol{X}_{bi})^{-1}\boldsymbol{X}_{bi}^H\boldsymbol{y}, \hat{\gamma}^{-1}(\boldsymbol{X}_{bi}^H\boldsymbol{X}_{bi})^{-1}\right),
\end{align}
where
\begin{align}
\hat{\gamma} &= \int \gamma b(\gamma) d\gamma  \nonumber \\
&= \frac{Z}{(\boldsymbol{y} - \boldsymbol{X}_{bi}\boldsymbol{c}^p)^H(\boldsymbol{y} - \boldsymbol{X}_{bi}{c}^p) + Tr\{\boldsymbol{X}_{bi}{V_c}^p\boldsymbol{X}_{bi}^H\}}. \label{eq:noiseprecision}
\end{align}
As the incoming message $m_{f_{c_{jb}}\rightarrow c_{jb}}(c_{jb})\propto \mathcal{CN}(c_{jb};\larrow{c_{jb}},\larrow{\nu}_{c_{jb}})$,  the belief of $\boldsymbol{c}$ is Gaussian with covariance matrix $\boldsymbol{V}_c^p$ and mean vector $\boldsymbol{c}^p$, which can be computed as
\begin{align}
\boldsymbol{V}_c^p =& (\boldsymbol{V}_c^{-1} +\hat{\gamma}\boldsymbol{X}_{bi}^H\boldsymbol{X}_{bi})^{-1},\label{eq:beliefcvar} \\
\boldsymbol{c}^p =& \boldsymbol{V}^p_c (\boldsymbol{V}_c^{-1}\larrow{\boldsymbol{c}} + \hat{\gamma}\boldsymbol{X}_{bi}^H\boldsymbol{y}),\label{eq:beliefcmean}
\end{align}
{where $\boldsymbol{V}_{\boldsymbol{c}}$ is a diagonal matrix with the diagonal elements given by $\{\larrow{\nu}_{c_{jb}}\}$ and $\larrow{\boldsymbol{c}}$ is a column vector that consists of $\{\larrow{c}_{jb}\}$. The computations of the mean $\larrow{c}_{jb}$ and the variance $\larrow{\nu}_{c_{jb}}$ are delayed to \eqref{eq:cpriormean} and \eqref{eq:cpriorvar}.}
Hence the outgoing message $m_{f_{\boldsymbol{y}}\rightarrow {c}_{jb}}(c_{jb})$ is also Gaussian, and can be expressed as \cite{GuoAConcise}
\begin{align}
  m_{f_{\boldsymbol{y}}\rightarrow {c}_{jb}}(c_{jb}) = \mathcal{CN}(c_{jb};\rarrow{c}_{jb},\rarrow{\nu}_{c_{jb}})
\end{align}
with
\begin{align}
\rarrow{\nu}_{c_{jb}} =& \left(\frac{1}{\nu_{c_{jb}}^p}  - \frac{1}{\larrow{\nu}_{c_{jb}}}\right)^{-1}, \label{eq:extrinsicvar}\\
\rarrow{c}_{jb} =& \rarrow{\nu}_{c_{jb}}\left(\frac{{c}_{jb}^p}{\nu_{c_{jb}}^p} - \frac{\larrow{c}_{jb}}{\larrow{\nu}_{c_{jb}}}\right),\label{eq:extrinsicmean}
\end{align}
where ${c}_{jb}^p$ is the $b$th element of the $j$th block of ${\boldsymbol{c}}^p$, and $\nu_{c_{jb}}^p$ is the $b$th element of the $j$th block of the vector that consists of the diagonal elements of $\boldsymbol{V}_{\boldsymbol{c}}^p$.

\subsection{Message Computations in Part \uppercase\expandafter{\romannumeral2}}

\subsubsection{Forward Message Passing} With the incoming message from Part \uppercase\expandafter{\romannumeral1} $n_{c_{jb}\rightarrow f_{c_{jb}}} (c_{jb}) = m_{f_{\boldsymbol{y}}\rightarrow {c}_{jb}} (c_{jb}) = \mathcal{CN}(c_{jb};\rarrow{c}_{jb}, \rarrow{\nu}_{c_{jb}})$ and the factor $f_{c_{jb}}(c_{jb}, h_j,g_{jb})=\delta(c_{jb}-h_jg_{jb})$, we can obtain an intermediate function node $\tilde{f}_{c_{jb}}(h_j, g_{jb})$ according to BP \cite{Kschischang2001}, i.e.,
\begin{align}
  \tilde{f}_{c_{jb}}(h_j, g_{jb}) =& \int f_{c_{jb}}(c_{jb}, h_j,g_{jb}) n_{c_{jb}\rightarrow f_{c_{jb}}}(c_{jb})d{c_{jb}} \nonumber \\
   =& \mathcal{CN}(h_jg_{jb};\rarrow{c}_{jb}, \rarrow{\nu}_{c_{jb}}).
\end{align}
With the local function $\tilde{f}_{c_{jb}}(h_j, g_{jb})$, we can compute the message from  $f_{c_{jb}}$ to $h_j$ by treating $g_{jb}$ as a constant, i.e.,
\begin{align}
m_{f_{c_{jb}}\rightarrow h_j} (h_j) = \mathcal{CN}(h_j; \rarrow{h}_{jb}, \rarrow{\nu}_{h_{jb}}),
\end{align}
where
\begin{align}
\rarrow{h}_{jb} =& \frac{\rarrow{c}_{jb}}{\hat{g}_{jb}}, \label{add1h} \\
 \rarrow{\nu}_{h_{jb}} =& \frac{\rarrow{\nu}_{c_{jb}}}{|\hat{g}_{jb}|^2} \label{eq:left2h}
\end{align}
with $\hat{g}_{jb}$ being the mean of the Gaussian belief of $g_{jb}$, which is computed in (\ref{eq:vbeliefmean}).
\rev{The product of the Gaussian messages $\{m_{f_{c_{jb}}\rightarrow h_j}(h_j), \forall b \}$ is still Gaussian \cite{GuoAConcise}, i.e.,}
\begin{align}
  q_j(h_j) = \prod_b m_{f_{c_{jb}}\rightarrow h_j} (h_j)\propto \mathcal{CN}(h_j;\hat{q}_j, \nu_{q_j}),
\end{align}
where
\begin{align}
  {\nu}_{q_j} =& \left(\sum_b\frac{1}{\rarrow{\nu}_{h_{jb}}}\right)^{-1}, \\
  \hat{q}_j =& {\nu}_{q_j}\sum_b\frac{\rarrow{h}_{jb}}{\rarrow{\nu}_{h_{jb}}}.
\end{align}
With the message $m_{f_{h_j}\rightarrow h_j} (h_j) \propto \mathcal{CN}(h_j;0,\widehat{\lambda}_j^{-1})$, which is given later in (\ref{eq:fg2g}), the belief $b(h_j)$ of  $h_j$ is obtained as
\begin{align}
b(h_j) = q_j(h_j)m_{f_{h_j}\rightarrow h_j}(h_j) \propto \mathcal{CN}(h_j;\hat{h}_j, \nu_{h_j}),
\end{align}
where
\begin{align}
  \hat{h}_j =& \frac{\hat{q}_j}{1+{\nu}_{q_j}\widehat{\lambda}_j}, \label{eq:gbeliefmean} \\
  \nu_{h_j} =& \left(\frac{1}{{\nu}_{q_j}} + \widehat{\lambda}_j\right)^{-1}. \label{eq:gbeliefvar}
\end{align}
The message $m_{f_{h_j}\rightarrow \lambda_j}(\lambda_j)$ is then computed by using the MF rule, i.e.,
\begin{align}
    m_{f_{h_j}\rightarrow \lambda_j}(\lambda_j) \propto& exp\left\{\int \ln f_{h_j}(h_j, \lambda_j)b(h_j)d{h_j}\right\}\nonumber \\ \propto & \lambda_jexp\{-\lambda_j(|\hat{h}_j|^2 + \nu_{h_j})\},
\end{align}
so the belief $b(\lambda_j)$ of the hyperparameter $\lambda_j$ is given as
\begin{align}
b(\lambda_j) =& m_{f_{h_j}\rightarrow \lambda_j}(\lambda_j)f_{\lambda_j}(\lambda_j) \nonumber \\ \propto& \lambda_j^{\epsilon_j} exp\left\{-\lambda_j(\eta_j+|\hat{h}_j|^2 + \nu_{h_j})\right\}.
\end{align}
The message from $f_{c_{jb}}$ to $g_{jb}$ is computed with the MF rule, i.e.,
\begin{align}
  m_{f_{c_{jb}}\rightarrow g_{jb}} (g_{jb}) \propto& exp\left\{\int \ln\tilde{f}_{c_{jb}}(h_j, g_{jb})b(h_j)d{h_{j}}\right\} \nonumber \\
=& \mathcal{CN}(g_{jb}; \rarrow{g}_{jb}, \rarrow{\nu}_{g_{jb}}),
\end{align}
where
\begin{align}
\rarrow{g}_{jb} = & \frac{\rarrow{c}_{jb}\hat{h}_{j}^*}{|\hat{h}_j|^2 + \nu_{h_j}}, \label{eq:rightvjbmean} \\
\rarrow{\nu}_{g_{jb}} = & \frac{\rarrow{\nu}_{c_{jb}}}{|\hat{h}_j|^2 + \nu_{h_j}}. \label{eq:rightvjbvar}
\end{align}
It is noted that the local function node $f_{g_jb}$ includes a nonlinear function $\Phi(q,\kappa_j)$, which makes the message computation about $\kappa_j$ intractable. \rev{Inspired by the extended Kalman filter \cite{extended} }, to solve this problem, $\Phi(q,\kappa_j)$ is linearized by using the first order Taylor expansion with the estimate of $\kappa_j$ in last iteration, i.e.,
\begin{align}
\Phi(q,\kappa_j) \approx \Phi(q, \hat{\kappa}_j^{'}) + \Phi'(q, \hat{\kappa}_j^{'})(\kappa_j - \hat{\kappa}_j^{'}), \label{eq:taylor}
\end{align}
with
\begin{align}
\Phi'(q, \hat{\kappa}_j^{'}) =&  \left(\frac{-j2\pi t}{MN}\right)\Phi(q, \hat{\kappa}_j^{'})+ \nonumber\\ & e^{-j2\pi\frac{t(d+\hat{\kappa}_j^{'})}{MN}} \frac{1}{N}\sum_{n=1}^{N-1}j\frac{2n\pi}{N}e^{j\frac{2n\pi}{N}(q + \hat{\kappa}_j^{'})},
\end{align}
where $\hat{\kappa}_j^{'}$ denotes the estimates of $\kappa_j$ in last iteration.} {Then, with the BP rule and the approximation in (\ref{eq:taylor}), the message $m_{f_{g_{jb}}\rightarrow \kappa_j} (\kappa_j)$ can be expressed as
\rev{
\begin{align}
	m_{f_{g_{jb}}\rightarrow \kappa_j}(\kappa_j) =&
	\int  f_{g_{jb}}(g_{jb},\kappa_j)m_{f_{c_{jb}}\rightarrow g_{jb}}(g_{jb}) d{g_{jb}}. \label{eq:msgfgjb2kappa}
\end{align}
Note that $\kappa_j$ is a real valued variable. To ensure that the message $m_{f_{g_{jb}}\rightarrow \kappa_j} (\kappa_j)$ is real, we rewrite the function $f_{g_{jb}}(g_{jb},\kappa_j)$ as
\begin{align}
&f_{g_{jb}}(g_{jb},\kappa_j)\nonumber\\ 
&=\delta\left(\mathcal{R}[g_{jb}] - \mathcal{R}[\Phi(Q,\kappa_j)]\right)\delta\left(\mathcal{I}[g_{jb}] - \mathcal{I}[\Phi(Q,\kappa_j)]\right),
\end{align}
where $Q=-\widehat{N}+b-1$. Then the message $m_{f_{g_{jb}}\rightarrow \kappa_j}(\kappa_j)$ in (\ref{eq:msgfgjb2kappa}) can be obtained as
\begin{align}
	m_{f_{g_{jb}}\rightarrow \kappa_j}(\kappa_j) \propto & \mathcal{N}(\kappa_j; \rarrow{\kappa}_{jb}^{\mathcal{R}},\rarrow{\nu}_{\kappa_{jb}}^{\mathcal{R}})\mathcal{N}(\kappa_j; \rarrow{\kappa}_{jb}^{\mathcal{I}},\rarrow{\nu}_{\kappa_{jb}}^{\mathcal{I}}) \nonumber \\
	\propto & \mathcal{N} (\kappa_j; \rarrow{\kappa}_{jb},\rarrow{\nu}_{\kappa_{jb}}),	
\end{align}
where
\begin{align}
	\rarrow{\kappa}_{jb}^{\mathcal{R}} =& \frac{\mathcal{R}[\rarrow{g}_{jb}] - \mathcal{R}[\Phi(Q, \hat{\kappa}_j^{'})] + \mathcal{R}[\Phi'(Q, \hat{\kappa}_j^{'})]\hat{\kappa}_j^{'}}{\mathcal{R}[\Phi'(Q, \hat{\kappa}_j^{'})]} , \\
	\rarrow{\nu}_{\kappa_{jb}}^{\mathcal{R}} =&  \frac{\rarrow{\nu}_{g_{jb}}}{2|\mathcal{R}[\Phi'(Q, \hat{\kappa}_j^{'})]|^2},\\
	\rarrow{\kappa}_{jb}^{\mathcal{I}} =& \frac{\mathcal{I}[\rarrow{g}_{jb}] - \mathcal{I}[\Phi(Q, \hat{\kappa}_j^{'})] + \mathcal{I}[\Phi'(Q, \hat{\kappa}_j^{'})]\hat{\kappa}_j^{'}}{\mathcal{I}[\Phi'(Q, \hat{\kappa}_j^{'})]} , \\
	\rarrow{\nu}_{\kappa_{jb}}^{\mathcal{I}} =&  \frac{\rarrow{\nu}_{g_{jb}}}{2|\mathcal{I}[\Phi'(Q, \hat{\kappa}_j^{'})]|^2},
\end{align}
and 
\begin{align}
	\rarrow{\nu}_{\kappa_{jb}} =& \left(\frac{1}{\rarrow{\nu}_{\kappa_{jb}}^{\mathcal{R}}} + \frac{1}{\rarrow{\nu}_{\kappa_{jb}}^{\mathcal{I}}}\right)^{-1},\label{eq:rightkappajbmean} \\
	\rarrow{\kappa}_{jb} =& \rarrow{\nu}_{\kappa_{jb}}\left(\frac{\rarrow{\kappa}_{jb}^{\mathcal{R}}}{\rarrow{\nu}_{\kappa_{jb}}^{\mathcal{R}}} + \frac{\rarrow{\kappa}_{jb}^{\mathcal{I}}}{\rarrow{\nu}_{\kappa_{jb}}^{\mathcal{I}}}\right).\label{eq:rightkappajbvar}
\end{align}
}
The prior of $\kappa_j$ is a uniform distribution over $[-0.5, 0.5]$, i.e., $-0.5 \leq \kappa_j \leq 0.5$. To simplify the computation of the belief $b(\kappa_j)$, we simply carry out the clipping operation, i.e.,
\begin{equation}
    b(\kappa_j)\propto\begin{cases}
    \mathcal{N}(\kappa_j; -0.5, 0); & \hat\kappa_j \leq -0.5 \\
    \mathcal{N}(\kappa_j; 0.5, 0); & \hat\kappa_j \geq 0.5 \\
    \mathcal{N}(\kappa_j; \hat{\kappa}_j, \nu_{\kappa_j}); & otherwise \\
    \end{cases} \label{eq:kappabelief}
\end{equation}
where
\begin{equation}
  \nu_{\kappa_j} = \left(\sum_b\frac{1}{\rarrow{\nu}_{\kappa_{jb}}}\right)^{-1},
  \end{equation}
  \begin{equation}
  \hat{\kappa}_j = \nu_{\kappa_j}\sum_b\frac{\rarrow{\kappa}_{jb}}{\rarrow{\nu}_{\kappa_{jb}}}.
\end{equation}

\subsubsection{Backward Message Passing} We firstly compute the message $m_{f_{h_j}\rightarrow h_j}(h_j)$ from the function node $f_{h_j}$ to $h_j$ by using the MF rule as follows
\begin{align}
m_{f_{h_j}\rightarrow h_j}(h_j) \propto & exp\left\{\int \ln f_{h_j}(h_j, \lambda_j)b(\lambda_j)d{\lambda_j}\right\} \nonumber \\
\propto & \mathcal{CN}(h_j; 0, \hat{\lambda}_j^{-1}), \label{eq:fg2g}
\end{align}
where
\rev{
\begin{align}
    \hat{\lambda}_j = \int \lambda_j b(\lambda_j)d\lambda_j = \frac{\epsilon_j+1}{\eta_j + |\hat{h}_j|^2+\nu_{h_j}}. \label{eq:lambdaj}
\end{align}
}
Then the message $n_{h_j\rightarrow f_{c_{jb}}}(h_j)$ from variable node $h_j$ to $f_{c_{jb}}$ is updated by the BP rule, i.e.,
\begin{align}
n_{h_j\rightarrow f_{c_{jb}}}(h_j) = \frac{b(h_j)}{m_{f_{c_{jb}} \rightarrow h_j}} \propto \mathcal{CN}(h_j;\larrow{h}_{jb}, \larrow{\nu}_{h_{jb}}),
\end{align}
where
\begin{align}
\larrow{\nu}_{h_{jb}} =& \left({\frac{1}{\nu_{h_j}}} - \frac{1}{\rarrow{\nu}_{h_{jb}}}\right)^{-1}, \label{add}
\end{align}
\begin{align}
\larrow{h}_{jb} =& \larrow{\nu}_{h_{jb}}\left(\frac{\hat{h}_j}{\nu_{h_j}} - \frac{\rarrow{h}_{jb}}{\rarrow{\nu}_{h_{jb}}}\right). \label{add1}
\end{align}
The backward message $n_{\kappa_j\rightarrow f_{g_{jb}}}(\kappa_j)$ is Gaussian with mean $\larrow{\kappa}_{jb}$ and variance $\larrow{\nu}_{\kappa_{jb}}$ and can be calculated as
\begin{align}
  \larrow{\nu}_{\kappa_{jb}} =& (1/\nu_{\kappa_j} - 1/\rarrow{\nu}_{\kappa_{jb}})^{-1}, \label{eq:leftkappajbvar}\\
  \larrow{\kappa}_{jb} =& \larrow{\nu}_{\kappa_{jb}} (\hat{\kappa}_j/\nu_{\kappa_j} - \rarrow{\kappa}_{jb}/\rarrow{\nu}_{\kappa_{jb}}).\label{eq:leftkappajbmean}
\end{align}
For the cases that $\hat{\kappa}_j=0.5$ or $-0.5$ and $\nu_{\kappa_j}=0$ in (\ref{eq:kappabelief}), we set
\begin{align}
  \larrow{\nu}_{\kappa_{jb}} = \nu_{\kappa_j}, \quad \larrow{\kappa}_{jb} = \hat{\kappa}_j.
\end{align}
Then the message $m_{f_{g_{jb}}\rightarrow g_{jb}}(g_{jb})$ is calculated by using the BP rule, i.e.,
\begin{align}
m_{f_{g_{jb}}\rightarrow g_{jb}} (g_{jb})=& \int {f_{g_{jb}}(g_{jb},\kappa_j)n_{\kappa_j\rightarrow f_{g_{jb}}}(\kappa_j)}d{\kappa_j} \nonumber \\ \propto & \mathcal{CN}(g_{jb};\larrow{g}_{jb},\larrow{\nu}_{g_{jb}}),
\end{align}
where
\begin{align}
    \larrow{g}_{jb} =& \Phi(Q, \hat{\kappa}_j^{'}) + \Phi'(Q, \hat{\kappa}_j^{'})(\larrow{\kappa}_{jb} - \hat{\kappa}_j^{'}), \label{eq:leftvjbmean}\\
    \larrow{\nu}_{g_{jb}}=& \larrow{\nu}_{\kappa_{jb}}|\Phi'(Q, \hat{\kappa}_j^{'})|^2. \label{eq:leftvjbvar}
\end{align}
Hence we can obtain the belief $b(g_{jb})\propto \mathcal{CN}(g_{jb};\hat{g}_{jb},\nu_{g_{jb}})$ with
\begin{equation}
  \nu_{g_{jb}} = \left(\frac{1}{\larrow{\nu}_{g_{jb}}} + \frac{1}{\rarrow{\nu}_{g_{jb}}}\right)^{-1}, \label{eq:vbeliefvar} \\
\end{equation}
\begin{equation}
  \hat{g}_{jb} = \nu_{g_{jb}}\left(\frac{\larrow{g}_{jb}}{\larrow{\nu}_{g_{jb}}} + \frac{\rarrow{g}_{jb}}{\rarrow{\nu}_{g_{jb}}}\right). \label{eq:vbeliefmean}
\end{equation}
Finally, by combining the incoming message $n_{g_{jb}\rightarrow f_{c_{jb}}}(g_{jb}) = m_{f_{g_{jb}}\rightarrow g_{jb}}(g_{jb})$ and $n_{h_j\rightarrow f_{c_{jb}}}(h_j)\propto \mathcal{CN}(h_j;\larrow{h}_{jb}, \larrow{\nu}_{h_{jb}})$, the message
$m_{f_{c_{jb}}\rightarrow c_{jb}}(c_{jb})$ is Gaussian with mean $\larrow{c}_{jb}$ and variance $\larrow{\nu}_{c_{jb}}$, which are computed as \cite{BIUTAMP}
\begin{align}
\larrow{c}_{jb}=&\larrow{h}_{jb}\larrow{g}_{jb}, \label{eq:cpriormean}\\
\larrow{\nu}_{c_{jb}} =& |\larrow{h}_{jb}|^2 + |\larrow{g}_{jb}|^2\larrow{\nu}_{h_{jb}} + \larrow{\nu}_{h_{jb}}. \label{eq:cpriorvar}
\end{align}
The message passing algorithm is summarized in Algorithm \ref{algorithm:ideallmmse}. \rev{The algorithm can be terminated when it reaches a maximum number of iteration or the difference between the estimates of parameters of two consecutive iterations is less than a threshold.}

\subsection{Extension to Rectangular Waveform}

The derivation of Algorithm 1 in the above is for OTFS with the bi-orthogonal waveform. Thanks to the iterative process of the algorithm, it can be readily extended for OTFS with the rectangular waveform. As shown in \eqref{eq:rectX}, the difference between $\boldsymbol{X}_{bi}$ and $\boldsymbol{X}_{rect}$ is that there is a Doppler shift-dependent factor for each of the elements in $\boldsymbol{X}_{rect}$, and the parameter $\kappa_n$ is unknown. Here we note that when $z$ and $n$ are given, the values of $l_z$ and $d_n$ are known.

This problem can be solved \rev{by taking advantage of} the iterative estimation strategy.
We can start the iterative process with initialization $\kappa_n=0$ \rev{and treat ${\boldsymbol{X}}_{rect}$ as a known matrix with the estimated Doppler shifts in last iteration plugged in (\ref{eq:rectX})} (here we abuse the use of the notation ${\boldsymbol{X}}_{rect}$, which is actually an estimate of the true ${\boldsymbol{X}}_{rect}$ with the estimate of $\kappa_n$).
Then the algorithm developed for the OTFS with the bi-orthogonal waveform can be used to recover $\boldsymbol{c}$, i.e., $\kappa_d$ is estimated, so that $\kappa_n$ can be found based on the corresponding $\kappa_d$. Then with the estimated $\kappa_n$, ${\boldsymbol{X}}_{rect}$ is updated for the next round iteration. Hence, we only need to add an extra step after step \ref{al1:kappaupdate} of Algorithm \ref{algorithm:ideallmmse} to update ${\boldsymbol{X}}_{rect}$ for OTFS with the rectangular waveform.
\begin{algorithm}
	\setstretch{1.25}
	\caption{Message Passing Algorithm for OTFS Channel Estimation}
	Initialize $\larrow{c}_{jb} = 0$, $\larrow{\nu}_{c_{jb}} = 1$, \rev{$\hat{\lambda}_j = 1$} for $\forall j,b$; \rev{$\hat{\gamma}=1$}, and $t=1$. 
	\\
	\textbf{Repeat}
	\begin{algorithmic}[1]
		\STATE update $\boldsymbol{V}_c^p$ and $\boldsymbol{c}^p$ with (\ref{eq:beliefcvar}) and (\ref{eq:beliefcmean}); \\
		\STATE update noise precision \rev{$\hat{\gamma}$} with (\ref{eq:noiseprecision});\\
		\STATE update the $\rarrow{c}_{jb}$ and $\rarrow{\nu}_{c_{jb}}$ with (\ref{eq:extrinsicmean}) and (\ref{eq:extrinsicvar});\\
		\STATE \rev{update $\rarrow{h}_{jb}$ and $\rarrow{\nu}_{h_{jb}}$ with (\ref{add1h}) and (\ref{eq:left2h});} \\ 		
		\STATE update the belief $b(h_j)$ of $h_j$ with (\ref{eq:gbeliefmean}) and (\ref{eq:gbeliefvar});\\
		\STATE update \rev{$\hat{\lambda}_j$} with (\ref{eq:lambdaj});\\
        \STATE update the belief $b(h_j)$ of $h_j$ with (\ref{eq:gbeliefmean}) and (\ref{eq:gbeliefvar});\\
        \STATE update $\larrow{h}_{jb}$ and $\larrow{\nu}_{h_{jb}}$ with (\ref{add1}) and (\ref{add});\\	
        \STATE update $\rarrow{g}_{jb}$ and $\rarrow{\nu}_{g_{jb}}$ with (\ref{eq:rightvjbmean}) and (\ref{eq:rightvjbvar}); \\
        \STATE update $\rarrow{\kappa}_{jb}$ and $\rarrow{\nu}_{\kappa_{jb}}$ with (\ref{eq:rightkappajbmean}) and (\ref{eq:rightkappajbvar}); \\       
        \STATE update the belief $b(\kappa_j)$ of $\kappa_j$ with (\ref{eq:kappabelief}); \label{al1:kappaupdate}\\
        	\rev{update ${\boldsymbol{X}}_{rect}$ with estimated Doppler shifts plugged in (\ref{eq:rectX}) in the case of rectangular waveform;}\\
        \STATE update $\larrow{\nu}_{\kappa_{jb}}$ and $\larrow{\kappa}_{jb}$ with (\ref{eq:leftkappajbvar}) and (\ref{eq:leftkappajbmean});\\
        \STATE update $\larrow{g}_{jb}$ and $\larrow{\nu}_{g_{jb}}$ with (\ref{eq:leftvjbvar}) and (\ref{eq:leftvjbmean});\\
        \STATE update the belief $b(g_{jb})$ of $g_{jb}$ with (\ref{eq:vbeliefvar}) and (\ref{eq:vbeliefmean});\\
        \STATE update $\larrow{c}_{jb}$ and $\larrow{\nu}_{c_{jb}}$ with (\ref{eq:cpriormean}) and (\ref{eq:cpriorvar});\\
		\STATE  $t=t+1$.
	\end{algorithmic}
	\textbf{Until terminated}
	\label{algorithm:ideallmmse}
\end{algorithm}

\section{Cramer-Rao Lower Bound} \label{sec:crlb}

 To evaluate the performance of the proposed algorithm, we derive the CRLB for the channel gain and Doppler shift estimation. We assume that the locations of the nonzero elements in $\boldsymbol{c}$ are known to facilitate the derivation of the CRLB. Hence, the CRLB derived in this section is a loose performance bound for the estimation.

 We firstly rewrite (\ref{eq:IdealFracY}) and (\ref{eq:RectFracYSimplified}) as
 \begin{align}
 y[k,l] = u[k,l] + \omega[k,l],
 \end{align}
 where
 \begin{align}
   u[k,l] = \sum_{i=1}^P\rev{\sum_{q=-\hat{N}}^{\hat{N}}}& h_if(q,\kappa_i)e^{-j2\pi \frac{l_i(k_i+\kappa_i)}{MN}} \nonumber \\
&\times x([k-k_i+q]_N,[l-l_i]_M)
 \end{align}
 for the bi-orthogonal waveform, and
\begin{align}
   u[k,l] = \sum_{i=1}^P\rev{\sum_{q=-\hat{N}}^{\hat{N}}}&h_if(q,\kappa_i)e^{j2\pi \frac{(l-l_i)(k_i+\kappa_i)}{MN}} \nonumber \\
 & \times x([k-k_i+q]_N,[l-l_i]_M)
 \end{align}
 for the rectangular waveform.
 In order to derive the CRLB, we define $\boldsymbol{\theta} = [h_1,h_p,\cdots,h_P,\kappa_1,\kappa_p,\cdots,\kappa_P]^T$. According to \cite{1993Fundamentals}, the CRLB of the $j$th element in $\boldsymbol{\theta}$ is the $j$th diagonal element of the inverse of the Fisher information
 matrix, i.e.,
 \begin{align}
   \theta_j^{CRLB} = [\boldsymbol{I}^{-1}(\boldsymbol{\theta})]_{jj},
 \end{align}
 where the Fisher information matrix $\boldsymbol{I}(\boldsymbol{\theta})$ has a size of $2P \times 2P$ with the $(i,j)$th element given as
 \begin{align}
 [\boldsymbol{I}(\boldsymbol{\theta})]_{ij} = -\mathbb{E}\left[\frac{\partial^2 \ln p(\boldsymbol{y};\boldsymbol{\theta}) }{\partial {\theta}_i\partial {\theta}_j}\right]
 \label{eq:fisherelement}
 \end{align}
 for $i=1,2,\cdots, 2P$, $j=1,2,\cdots, 2P$, and the expectation is taken with respect to $p(\boldsymbol{y};\boldsymbol{\theta})$. The logarithm of the likelihood function $\ln p(\boldsymbol{y};\boldsymbol{\theta})$ can be expressed as
\begin{align}
\ln p(\boldsymbol{y};\boldsymbol{\theta}) = -Z\ln(\pi\gamma^{-1}) - \gamma\sum_{z=0}^{Z-1}|y_z-u_z|^2,
\end{align}
where $y_z$ denotes the $z$th element in $\boldsymbol{y}$ and $u_z$ denotes the corresponding $u[k,l]$. Then, (\ref{eq:fisherelement}) can be rewritten as
\begin{align}
 [\boldsymbol{I}(\boldsymbol{\theta})]_{ij} = \gamma\sum_{z=0}^{Z-1}\left[-\frac{\partial u_z}{\partial {\theta}_i}\frac{\partial u_z^*}{\partial {\theta}_j} - \frac{\partial u_z^*}{\partial {\theta}_i}\frac{\partial u_z}{\partial {\theta}_j} \right].
\end{align}

For simplicity, we define $k' \triangleq [k-k_p+q]_N$ and $l' \triangleq [l-l_p]_M$.
When $1\leq p \leq P$ ($p$ represents $i$ or $j$), the derivative $\frac{\partial{u}_z}{\partial {\theta}_p}$ is about the channel gain $h_p$. For the bi-orthogonal waveform, we have
\begin{align}
    \frac{\partial{u}_z}{\partial h_p} = \sum_{q=-\widehat{N}}^{\widehat{N}}&f(q,\kappa_p)e^{-j2\pi \frac{l_p(k_p+\kappa_p)}{MN}} x[k',l'],
\end{align}
while for the rectangular waveform we have
\begin{align}
    \frac{\partial{u}_z}{\partial h_p} = \sum_{q=-\widehat{N}}^{\widehat{N}}&f(q,\kappa_p)e^{j2\pi \frac{(l-l_p)(k_p+\kappa_p)}{MN}} x[k',l'].
\end{align}
When $P < p' \leq 2P$ ($p'$ represents $i$ or $j$), the derivative $\frac{\partial{u}_z}{\partial {\theta}_{p'}}$ is about $\kappa_p$, where $p=p'-P$. For the bi-orthogonal waveform, we have
\begin{align}
 \frac{\partial{u}_z}{\partial \kappa_p} =& \sum_{q=-\widehat{N}}^{\widehat{N}}h_p \biggl[\left(\frac{1}{N}\sum_{n=1}^{N-1}j\frac{2n\pi}{N}e^{j\frac{2n\pi}{N}(q + \kappa_p)}\right)e^{-j2\pi \frac{l_p(k_p+\kappa_p)}{MN}} \nonumber \\ +& f(q,\kappa_p)\left(e^{-j2\pi \frac{l_p(k_i+\kappa_p)}{MN}}\frac{-j2\pi l_p}{MN}\right)\biggl]  x[k',l'],
\end{align}
while for the rectangular waveform
\begin{align}
 \frac{\partial{u}_z}{\partial \kappa_p} =& \sum_{q=-\widehat{N}}^{\widehat{N}}h_p \biggl[\left(\frac{1}{N}\sum_{n=1}^{N-1}j\frac{2n\pi}{N}e^{j\frac{2n\pi}{N}(q + \kappa_p)}\right)e^{j2\pi \frac{(l-l_p)(k_p+\kappa_p)}{MN}} \nonumber \\ +& f(q,\kappa_p)\left(e^{j2\pi \frac{(l-l_p)(k_p+\kappa_p)}{MN}}\frac{j2\pi (l-l_p)}{MN}\right)\biggl]  x[k',l'].
\end{align}

To evaluate the performance of the proposed algorithm in terms of the normalized mean squared error (NMSE), we use the average normalized CRLB for the estimation of channel gains and fractional Doppler shifts, which are defined as
\begin{align}
  \hat{\boldsymbol{h}}_{bound} =& \frac{\sum_{j=1}^{P}{\theta}_j^{CRLB}}{||\boldsymbol{h}_{\boldsymbol{\theta}}||_2^2},
\end{align}
\begin{align}
  \hat{\boldsymbol{\kappa}}_{bound} =& \frac{\sum_{j=P+1}^{2P}{\theta}_j^{CRLB}}{||\boldsymbol{\kappa}_{\boldsymbol{\theta}}||_2^2},
\end{align}
where $\boldsymbol{h}_{\boldsymbol{\theta}}=(h_1, h_2,\cdots, h_P)^T$ and $\boldsymbol{\kappa}_{\boldsymbol{\theta}}=(\kappa_1, \kappa_2,\cdots, \kappa_P)^T$.

\section{Simulation Results} \label{sec:simulation}

In this section, we evaluate the performance of the proposed message passing algorithm in terms of the NMSE of the estimated  channel parameters $\hat{\boldsymbol{h}}$ and $\hat{\boldsymbol{\kappa}}$, and the reconstructed channel matrices $\widehat{\boldsymbol{H}}_{bi}$ and $\widehat{\boldsymbol{H}}_{rect}$. The NMSE is defined as
\rev{
\begin{align}
 NMSE(\boldsymbol{x})=\frac{1/L\sum_{l=1}^{L}||\hat{\boldsymbol{x}}_l - \boldsymbol{x}||_2^2}{||\boldsymbol{x}||_2^2},
\end{align}
where $\hat{\boldsymbol{x}}_l$ denotes the estimate of a variable $\boldsymbol{x}$ in the $l$th trial, and $L$ is the number of trials.}
In addition, the bit error rate (BER) of data detection based on the reconstructed channel matrix is also evaluated. We set $M = 128$ and $N=32$, i.e., there are $32$ time slots and $128$ subcarriers in the TF domain. The carrier frequency is 3 GHz, the subcarrier spacing is 2 kHz, and the quadrature phase shift keying (QPSK) is adopted for modulation. The
velocity of the mobile user is set to be $v = 120 km/h$, leading to a maximum Doppler frequency shift index $k_{max}=4$. We assume that the maximum delay index {$l_{{max}}$ = 10}, the Doppler index of the $i$th path is uniformly distributed over $[-k_{max}, k_{max}]$, and the delay index is uniformly distributed over $[1, l_{max}]$, excluding the first path ($\l_i=0$). As mentioned before, the fractional Doppler $\kappa_i$ has a uniform distribution over $[-0.5, 0.5]$. The channel path gains $\{h_j\}$ are independently drawn from the complex Gaussian distribution $\mathcal{N}(0, 1/P)$. Furthermore, we define the SNR of pilot and data as
\rev{
\begin{align}
    {\rm SNRp}=&10\log10\left(\frac{\mathcal{P}_p}{\gamma^{-1}}\right), \\
    {\rm SNRd}=&10\log10\left(\frac{\mathbb{E}|x_d|^2}{\gamma^{-1}}\right),
\end{align}
respectively, where $\mathcal{P}_p$ denotes the average power of pilot symbols.}

\subsection{NMSE performance comparison}
We examine the NMSE performance for the estimation of ${\boldsymbol{h}}$, ${\boldsymbol{\kappa}}$ and ${\boldsymbol{H}}$ for both the bi-orthogonal and the rectangular waveforms. For comparison, we also include the threshold based channel estimation method in {\cite{pilotref}} for the case of the bi-orthogonal waveform (as it is not clear how to perform the channel estimation in the case of the rectangular waveform with fractional Doppler shifts). The CRLB derived in the previous section is also shown for reference.

\begin{figure}[t]
	\centering
	\includegraphics[width=0.9\columnwidth]{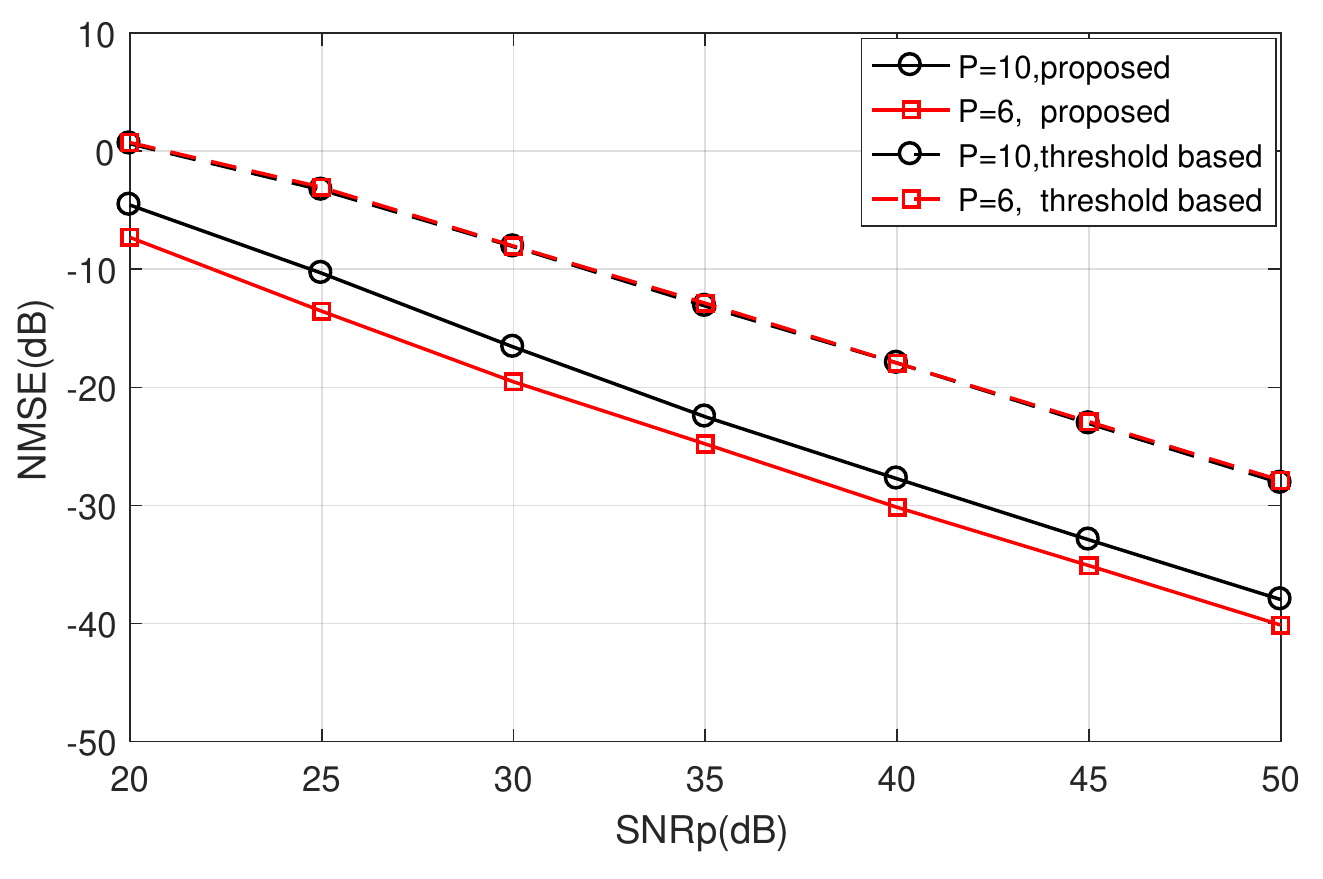}
	\caption{NMSE of $\widehat{\boldsymbol{H}}$ of the proposed algorithm and the threshold based method in \cite{pilotref}.} \label{fig:H_compare}
\end{figure}

\begin{table}[htb]
	\centering
	\caption{NMSE and PAPR comparisons: 1 pilot symbol versus 10 pilot symbols with the same power budget.}\label{tab:powerbudget} %
	\begin{tabular}{@{}ccccc}
		\midrule[1.4pt]
		\multirow{2}*{SNRp} & \multicolumn{3}{c}{NMSE} & \multirow{2}*{PAPR}\\
		\cmidrule(lr){2-4}
		~ & $\hat{\boldsymbol{h}}$ & $\hat{\boldsymbol{\kappa}}$ &  $\widehat{\boldsymbol{H}}$\\
		\midrule[1.4pt]
		50dB \\ 1 pilot ~ & -37.95 dB & -34.48 dB & -40.17 dB & 18.76 dB \\
		\midrule
		40dB \\ 10 pilots & -36.95 dB & -32.10 dB & -39.19 dB & 10.41 dB\\
		\midrule[1.4pt]
	\end{tabular}
\end{table}

\begin{figure}[t]
	\centering
	\includegraphics[width=0.9\columnwidth]{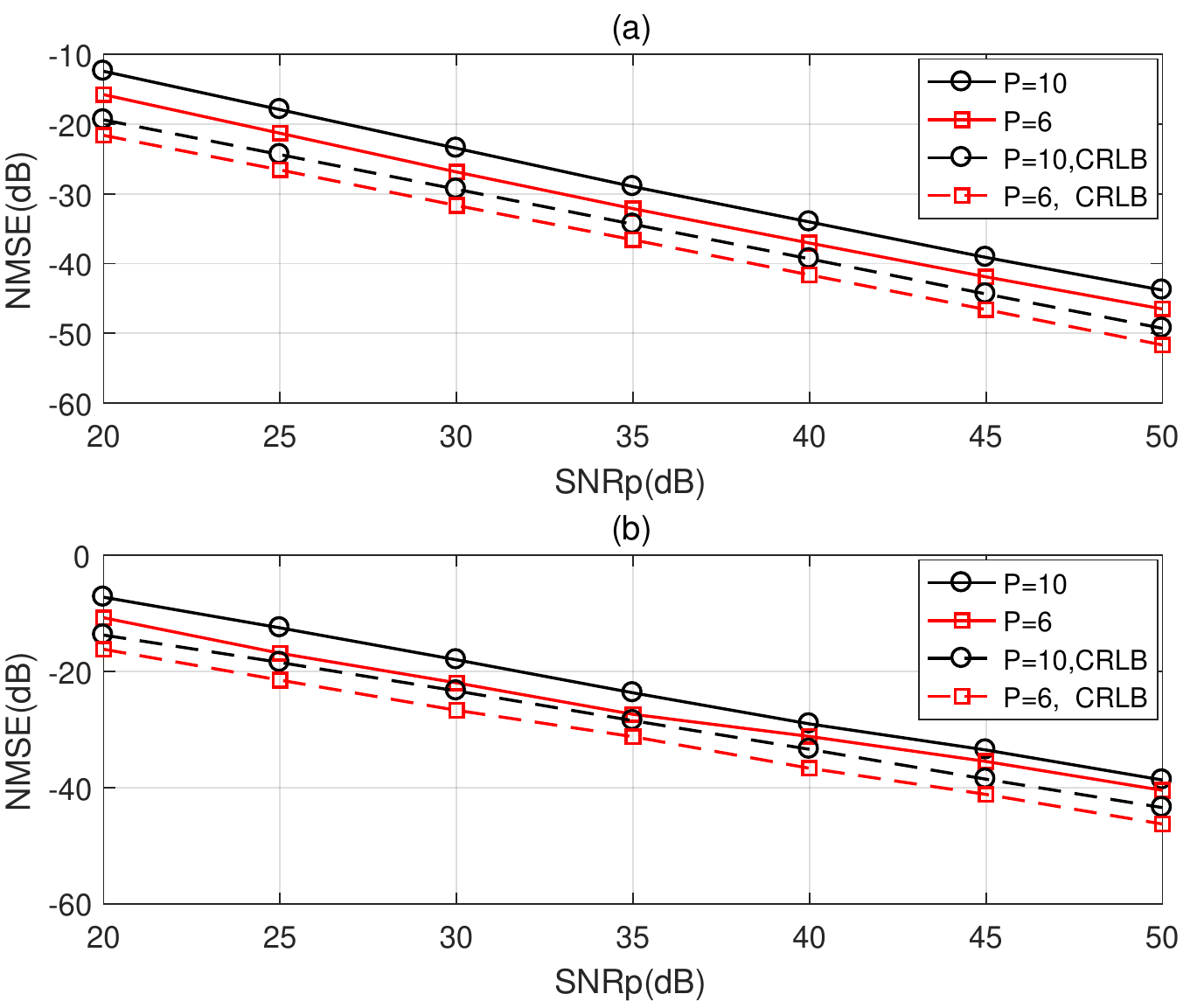}
	\caption{NMSE comparison for the bi-orthogonal waveform with different $P$: (a) NMSE of $\hat{\boldsymbol{h}}$; (b) NMSE of $\hat{\boldsymbol{\kappa}}$.} \label{fig:bi_nmse_pilot10}
\end{figure}
\begin{figure}[t]
	\centering
	\includegraphics[width=0.9\columnwidth]{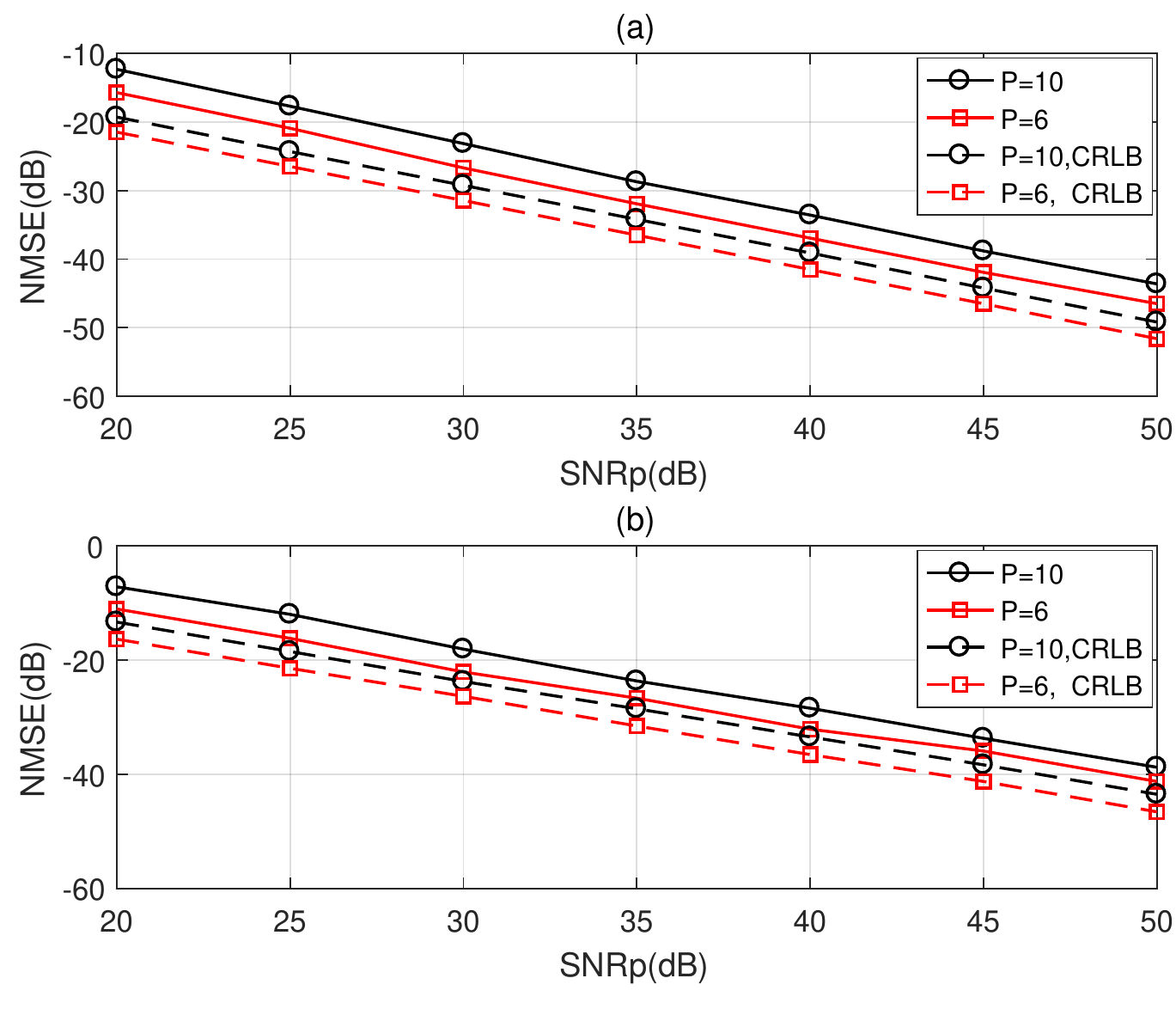}
	\caption{NMSE comparison for the rectangular waveform with different $P$: (a) NMSE of $\hat{\boldsymbol{h}}$; (b) NMSE of $\hat{\boldsymbol{\kappa}}$.} \label{fig:rect_nmse_pilot10}
\end{figure}

\begin{figure}[t]
	\centering
	\includegraphics[width=0.9\columnwidth]{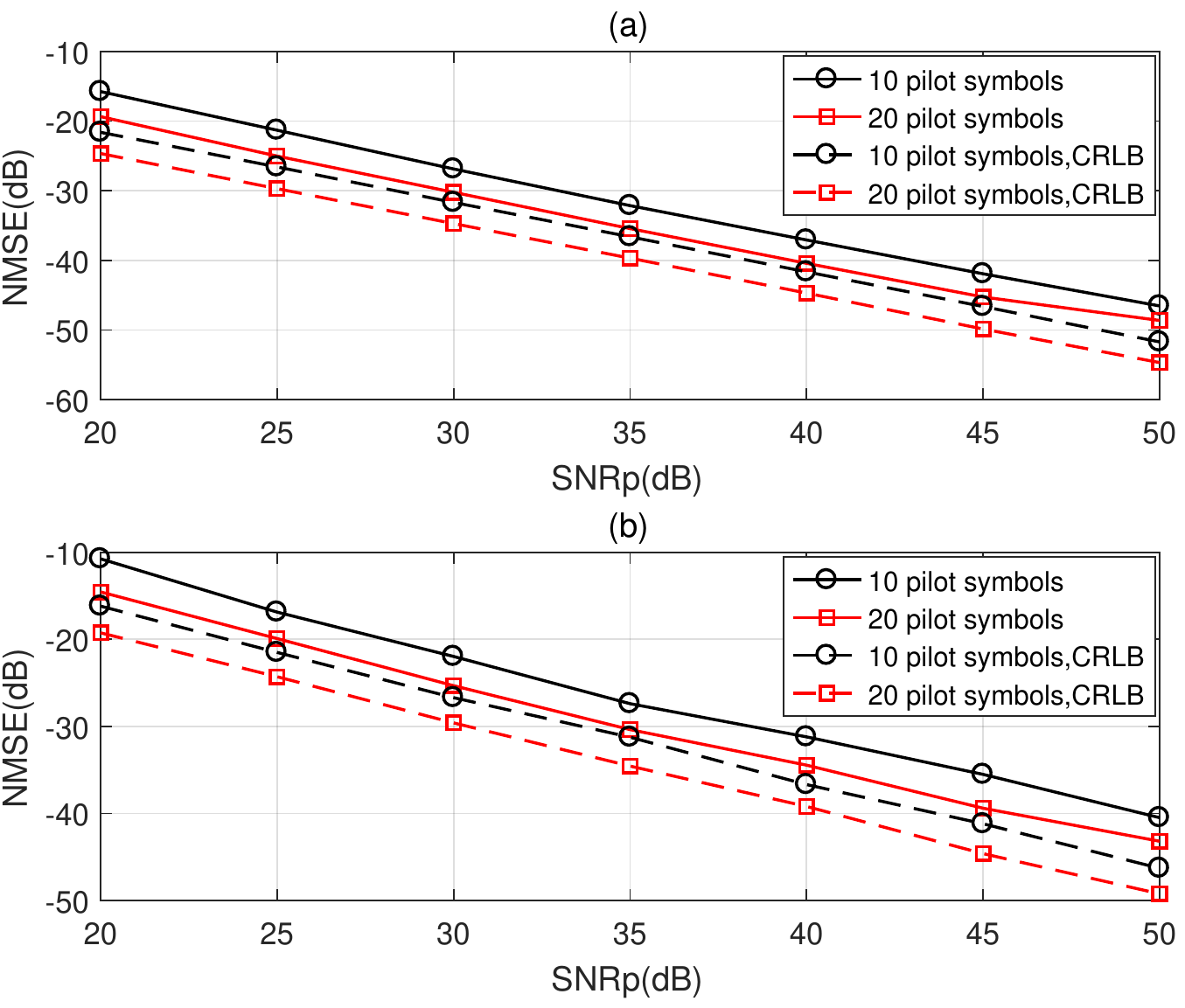}
	\caption{NMSE comparison for the bi-orthogonal waveform with different number of pilot symbols: (a) NMSE of $\hat{\boldsymbol{h}}$; (b) NMSE of $\hat{\boldsymbol{\kappa}}$.} \label{fig:bi_nmse_p6}
\end{figure}
\begin{figure}[t]
	\centering
	\includegraphics[width=0.9\columnwidth]{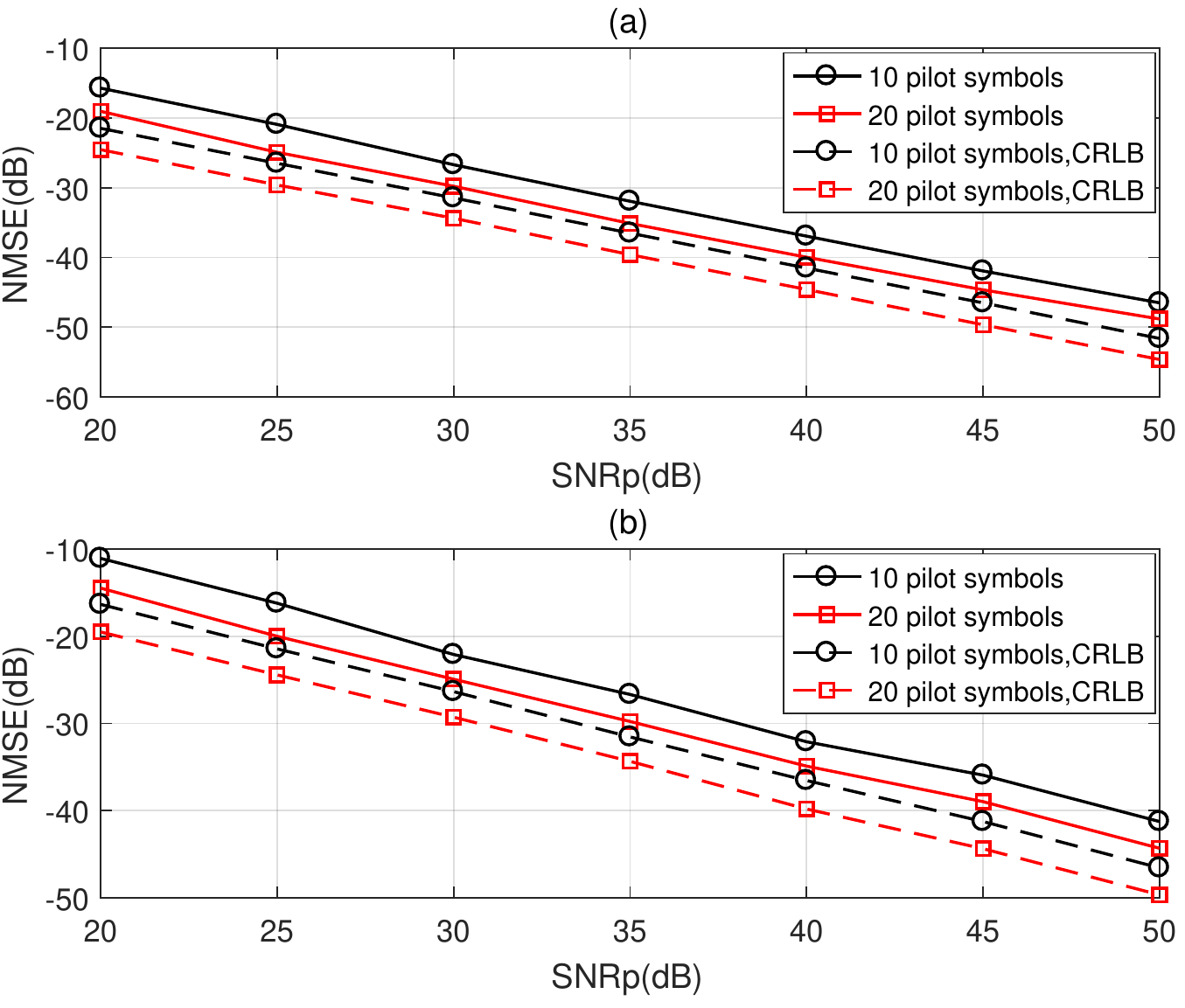}
	\caption{NMSE comparison for the rectangular waveform with different number of pilot symbols: (a) NMSE of $\hat{\boldsymbol{h}}$; (b) NMSE of $\hat{\boldsymbol{\kappa}}$.} \label{fig:rect_nmse_p6}
\end{figure}

In Fig. \ref{fig:H_compare}, we compare the NMSE performance of the reconstructed channel matrix $\widehat{\boldsymbol{H}}$ between our proposed algorithm and the threshold based method in \cite{pilotref}. Only a single pilot symbol is used and the number of paths $P=6$ and $10$ are considered. Moreover, with the recommendation in \cite{pilotref}, the threshold is set to be $3\sigma_p$ for the threshold based method, where $\sigma_p = (10^{{\rm SNRp}/10})^{-0.5}$. Note that the NMSE of $\hat{\boldsymbol{h}}$ and $\hat{\boldsymbol{\kappa}}$ cannot be compared as the threshold based method do not provide these estimates.
It can be seen that the proposed algorithm significantly outperforms the threshold based method in all the cases. The reason is that the threshold based method does not estimate the channel gains and fractional Doppler shifts, and it is equivalent to the estimate of the vector $\boldsymbol{c}$ without considering its structure. The results indicate that exploiting the structure of the vector $\boldsymbol{c}$ is crucial to improving the estimation performance.

{In Tab. \ref{tab:powerbudget}, we compare the PAPR and NMSE performance of $\hat{\boldsymbol{h}}$, $\hat{\boldsymbol{\kappa}}$ and $\widehat{\boldsymbol{H}}$ by using our proposed algorithm in two cases: a single pilot symbol and 10 pilot symbols. We assume a same power budget, so a single pilot symbol with SNRp = 50dB corresponds to 10 pilot symbols with SNRp = 40dB. From the table we can see that, the NMSE performance of using 10 pilot symbols is very close to that of using a single pilot symbol, but the PAPR with 10 pilot symbols is only 10.41 dB, in contrast to 18.76 dB for the case of using a single pilot symbol.}

The NMSE performances of the algorithm for different SNRp and number of paths $P$ are shown in Fig. \ref{fig:bi_nmse_pilot10} and Fig. \ref{fig:rect_nmse_pilot10}, where the number of pilot symbols is set to be 10. The performance of the threshold based method is not included as it only works with a single pilot symbol.
It can be seen that a smaller $P$ leads to a better performance as the vector $\boldsymbol{c}$ is more sparse and the number of parameters to be estimated is smaller. Comparing them to the CRLB, we can see that the NMSE performance of the proposed algorithm is fairly good as the CRLB is a loose lower bound, which is derived with the assumption that the locations and the number of the paths are known. In Fig. \ref{fig:bi_nmse_p6} and Fig. \ref{fig:rect_nmse_p6}, we show the performance of the proposed method with different numbers of pilot symbols. As expected, with more pilot symbols, the proposed algorithm delivers better performance.

\subsection{BER performance comparison}

\begin{figure}[htbp]
\centering
\subfigure[$P=6$; SNRp = 35dB; 1 pilot symbol]{
\centering
\includegraphics[width=0.4\columnwidth]{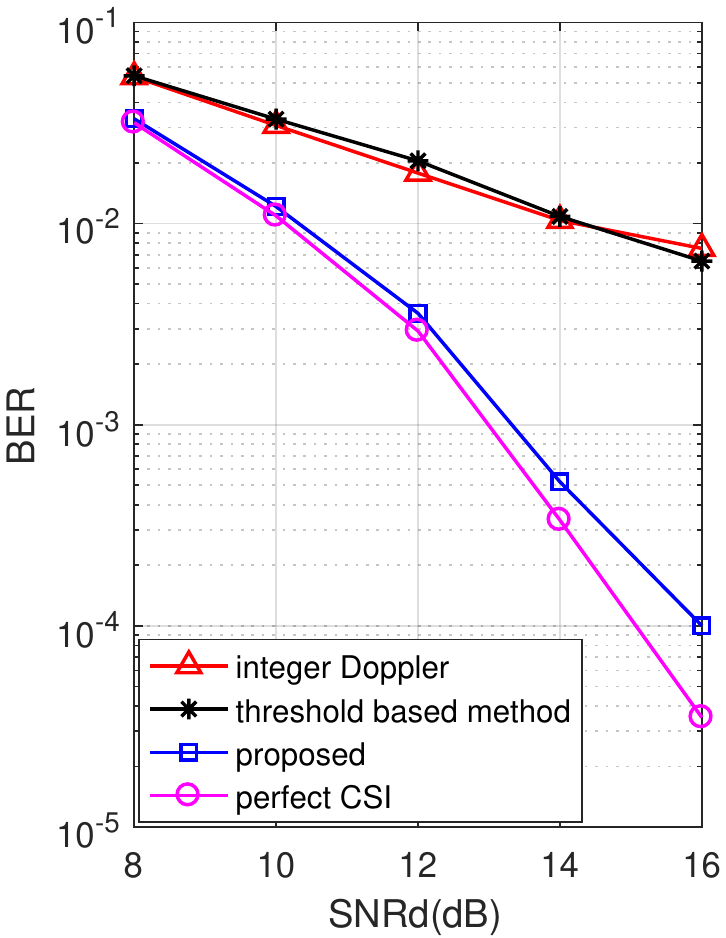} \label{fig:ber_p6_snrp40_pilot1}
}
\subfigure[$P=6$; SNRp = 40dB; 1 pilot symbol]{
\centering
\includegraphics[width=0.4\columnwidth]{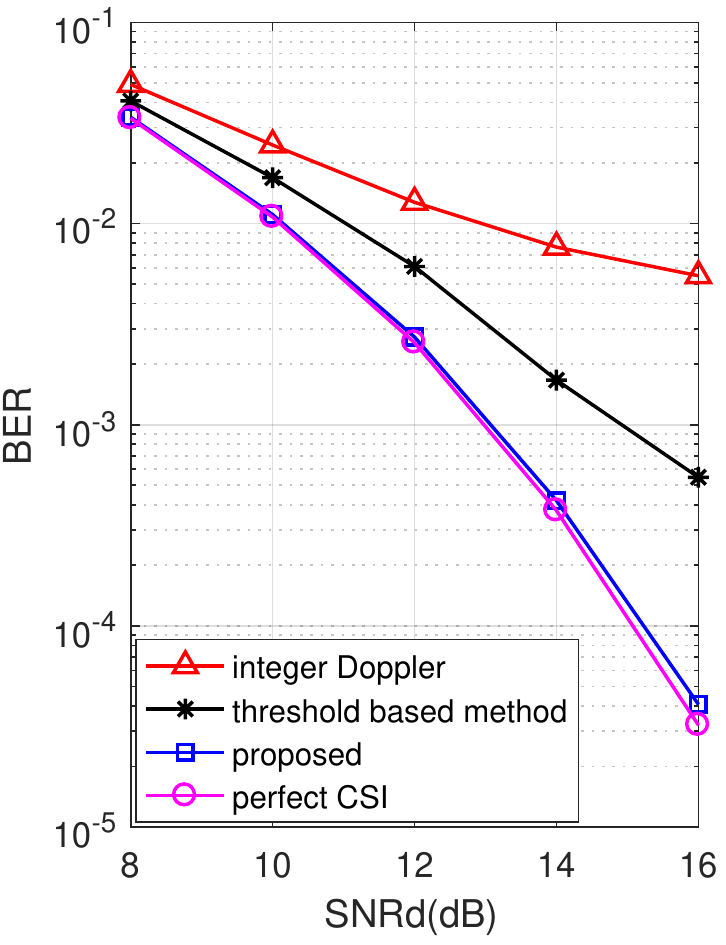} \label{fig:ber_p6_snrp45_pilot1}
}
\subfigure[$P=10$; SNRp = 40dB; 1 pilot symbol]{
\centering
\includegraphics[width=0.4\columnwidth]{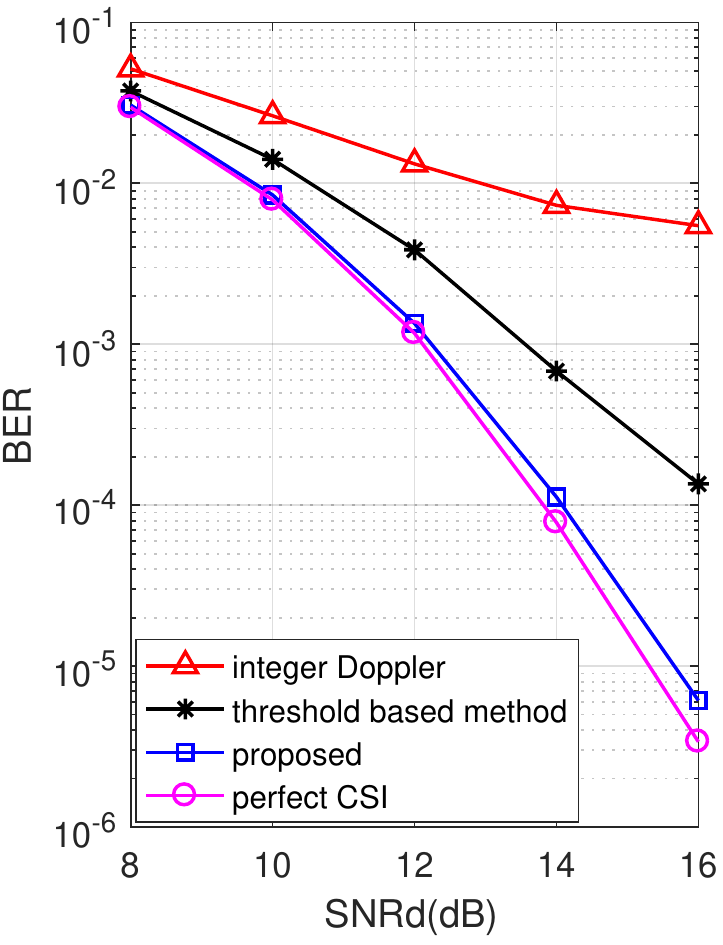} \label{fig:ber_p10_snrp40_pilot1}
}
\subfigure[$P=10$; SNRp = 40dB; 10 pilot symbols]{
\centering
\includegraphics[width=0.4\columnwidth]{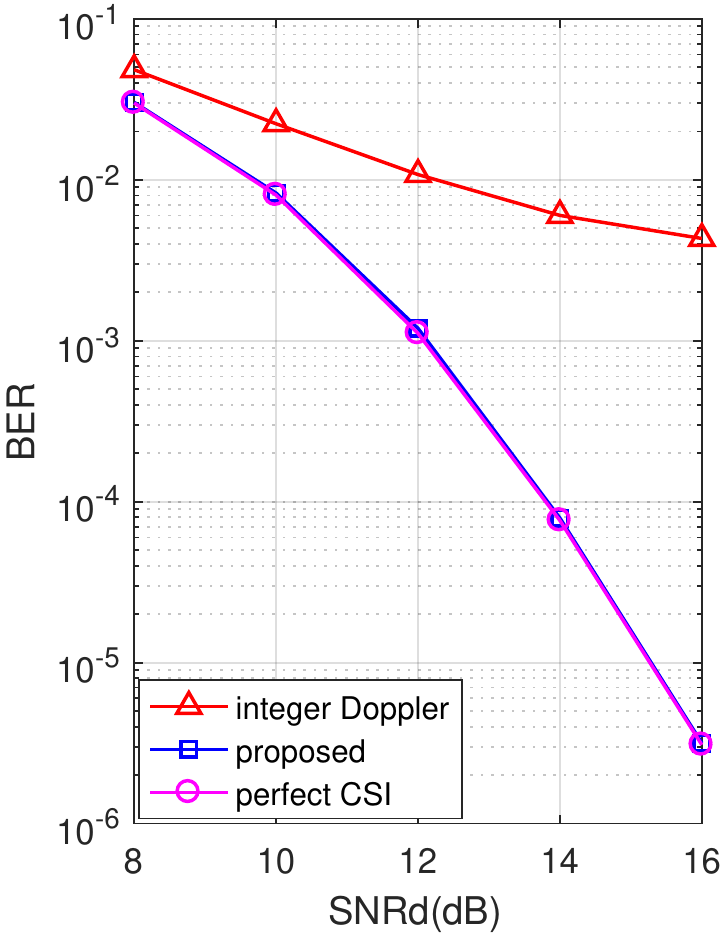} \label{fig:ber_p6_snrp40_pilot10}
}
\caption{BER performance comparison versus SNRp.} \label{fig:ber_comparison}
\end{figure}

We compare the BER performance of the OTFS system with the UTAMP based detector proposed in \cite{2020Iterative}. Assume that the bi-orthgonal waveform is used so that we can compare the performance of the system with the threshold based channel estimation method \cite{pilotref}. {As a benchmark, we also show the BER performance of the OTFS system with perfect channel matrix. In addition, in order to demonstrate the significance of considering fractional Doppler shifts, we include the performance of the OTFS system with the assumption of integer Doppler shifts (although fractional Doppler shifts exist in the generation of the OTFS channels).} The results are shown in Fig. \ref{fig:ber_comparison}. It can be seen that, when the fractional Doppler shifts are ignored, the system performance is degraded severely due to the modelling errors. In addition, in the case of single pilot symbol, the system with the proposed channel estimation algorithm delivers much better BER performance than that with the threshold based channel estimation method. \rev{Meanwhile}, the performance of the system with our proposed channel estimation algorithm can be very close to that of the system with perfect channel matrix.  By comparing Fig. \ref{fig:ber_p6_snrp45_pilot1} \rev{with} Fig. \ref{fig:ber_p10_snrp40_pilot1}, we can see that a larger $P$ leads to a better BER performance because more diversity gain can be achieved.

\section{Conclusions} \label{sec:conclusion}

In this paper, we have addressed the issue of the estimation of OTFS channels with fractional Doppler shifts, where both the bi-orthogonal waveform and the rectangular waveform are considered. The estimation is formulated as a sparse structured signal recovery problem \rev{and a Bayesian treatment is investigated}. With a factor graph representation of the problem, we have derived an message passing based algorithm to estimate the channel gains and fractional Doppler shifts. The CRLB has also been derived to evaluate the estimation performance of the proposed algorithm. It has been shown that the proposed algorithm significantly outperforms the existing algorithm, and it is able to work with multiple pilot symbols to achieve significant PAPR reduction.

\section*{Acknowledgment}
The authors would like to thank Prof. Jinhong Yuan at the University of New South Wales for his valuable comments and suggestions.

\bibliographystyle{IEEEtran}
\bibliography{IEEEabrv,bibliography}

\begin{thebibliography}{10}
\providecommand{\url}[1]{#1}
\csname url@samestyle\endcsname
\providecommand{\newblock}{\relax}
\providecommand{\bibinfo}[2]{#2}
\providecommand{\BIBentrySTDinterwordspacing}{\spaceskip=0pt\relax}
\providecommand{\BIBentryALTinterwordstretchfactor}{4}
\providecommand{\BIBentryALTinterwordspacing}{\spaceskip=\fontdimen2\font plus
\BIBentryALTinterwordstretchfactor\fontdimen3\font minus
  \fontdimen4\font\relax}
\providecommand{\BIBforeignlanguage}[2]{{%
\expandafter\ifx\csname l@#1\endcsname\relax
\typeout{** WARNING: IEEEtran.bst: No hyphenation pattern has been}%
\typeout{** loaded for the language `#1'. Using the pattern for}%
\typeout{** the default language instead.}%
\else
\language=\csname l@#1\endcsname
\fi
#2}}
\providecommand{\BIBdecl}{\relax}
\BIBdecl

\bibitem{Hadani2017}
R.~Hadani, S.~Rakib, M.~Tsatsanis, A.~Monk, and R.~Calderbank, ``{Orthogonal
  Time Frequency Space Modulation},'' in \emph{2017 IEEE Wireless
  Communications and Networking Conference (WCNC)}, 2017.

\bibitem{Raviteja2018}
P.~Raviteja, P.~K. T., H.~Yi, and V.~Emanuele, ``{Interference Cancellation and
  Iterative Detection for Orthogonal Time Frequency Space Modulation},''
  \emph{IEEE Transactions on Wireless Communications}, vol.~17, no.~10, pp.
  6501--6515, 2018.

\bibitem{OAP}
R.~Hadani and A.~Monk, ``{OTFS: A New Generation of Modulation Addressing the
  Challenges of 5G},'' \emph{ArXiv}, vol. abs/1802.02623, 2018.

\bibitem{2020Iterative}
Z.~Yuan, F.~Liu, W.~Yuan, Q.~Guo, Z.~Wang, and J.~Yuan, ``{Iterative Detection
  for Orthogonal Time Frequency Space Modulation with Unitary pproximate
  Message Passing },'' \emph{ArXiv}, vol. abs/2008.06688, 2020.

\bibitem{Raviteja2019}
P.~{Raviteja}, E.~{Viterbo}, and Y.~{Hong}, ``{OTFS Performance on Static
  Multipath Channels},'' \emph{IEEE Wireless Communications Letters}, vol.~8,
  no.~3, pp. 745--748, June 2019.

\bibitem{OTFSHighDopplerChannel}
K.~R. {Murali} and A.~{Chockalingam}, ``{On OTFS Modulation for High-Doppler
  Fading Channels},'' in \emph{2018 Information Theory and Applications
  Workshop (ITA)}, 2018, pp. 1--10.

\bibitem{performance2020}
S.~{Li}, J.~{Yuan}, W.~{Yuan}, Z.~{Wei}, B.~{Bai}, and D.~W.~K. {Ng},
  ``{Performance Analysis of Coded OTFS Systems over High-Mobility Channels},''
  \emph{arXiv e-prints}, p. arXiv:2010.13008, Oct. 2020.

\bibitem{LiA2017}
\BIBentryALTinterwordspacing
L.~Li, H.~Wei, Y.~Huang, Y.~Yao, W.~Ling, G.~Chen, P.~Li, and Y.~Cai, ``{A
  Simple Two-stage Equalizer With Simplified Orthogonal Time Frequency Space
  Modulation Over Rapidly Time-varying Channels},'' \emph{CoRR}, vol.
  abs/1709.02505, 2017. [Online]. Available:
  \url{http://arxiv.org/abs/1709.02505}
\BIBentrySTDinterwordspacing

\bibitem{Raviteja2019Practical}
P.~{Raviteja}, Y.~{Hong}, E.~{Viterbo}, and E.~{Biglieri}, ``Practical
  pulse-shaping waveforms for reduced-cyclic-prefix {OTFS},'' \emph{IEEE
  Transactions on Vehicular Technology}, vol.~68, no.~1, pp. 957--961, 2019.

\bibitem{zemen2017}
\BIBentryALTinterwordspacing
T.~Zemen, M.~Hofer, and D.~Loeschenbrand, ``{Low-Complexity Equalization for
  Orthogonal Time and Frequency Signaling (OTFS)},'' 2017. [Online]. Available:
  \url{http://arxiv.org/abs/1710.09916}
\BIBentrySTDinterwordspacing

\bibitem{otfslmmserecv}
S.~{Tiwari}, S.~S. {Das}, and V.~{Rangamgari}, ``{Low complexity LMMSE Receiver
  for OTFS},'' \emph{IEEE Communications Letters}, vol.~23, no.~12, pp.
  2205--2209, 2019.

\bibitem{fundWC}
D.~Tse and P.~Viswanath, \emph{Fundamentals of Wireless Communication}.\hskip
  1em plus 0.5em minus 0.4em\relax USA: Cambridge University Press, 2005.

\bibitem{pilotref}
P.~{Raviteja}, K.~T. {Phan}, and Y.~{Hong}, ``{Embedded Pilot-Aided Channel
  Estimation for OTFS in Delay–Doppler Channels},'' \emph{IEEE Transactions
  on Vehicular Technology}, vol.~68, no.~5, pp. 4906--4917, 2019.

\bibitem{MIMOOTFSDETECTIONEST}
M.~{Kollengode Ramachandran} and A.~{Chockalingam}, ``{MIMO-OTFS in
  High-Doppler Fading Channels: Signal Detection and Channel Estimation},'' in
  \emph{2018 IEEE Global Communications Conference (GLOBECOM)}, 2018, pp.
  206--212.

\bibitem{chestmassivemimo}
W.~{Shen}, L.~{Dai}, J.~{An}, P.~{Fan}, and R.~W. {Heath}, ``{Channel
  Estimation for Orthogonal Time Frequency Space (OTFS) Massive MIMO},''
  \emph{IEEE Transactions on Signal Processing}, vol.~67, no.~16, pp.
  4204--4217, 2019.

\bibitem{uplinkotfsmassivemimo}
Y.~{Liu}, S.~{Zhang}, F.~{Gao}, J.~{Ma}, and X.~{Wang}, ``{Uplink-Aided High
  Mobility Downlink Channel Estimation Over Massive MIMO-OTFS System},''
  \emph{IEEE Journal on Selected Areas in Communications}, vol.~38, no.~9, pp.
  1994--2009, 2020.

\bibitem{Monk2016OTFSO}
\BIBentryALTinterwordspacing
A.~Monk, R.~Hadani, M.~Tsatsanis, and S.~Rakib, ``{OTFS - Orthogonal Time
  Frequency Space},'' \emph{CoRR}, vol. abs/1608.02993, 2016. [Online].
  Available: \url{http://arxiv.org/abs/1608.02993}
\BIBentrySTDinterwordspacing

\bibitem{Kschischang2001}
{F.R. Kschischang and B.J. Frey and H.-A. Loeliger}, ``Factor graphs and the
  sum-product algorithm,'' \emph{{IEEE} Trans. Inform. Theory}, vol.~47, no.~2,
  pp. 498--519, Feb. 2001.

\bibitem{Tipping2001Sparse}
M.~E. Tipping, ``{Sparse Bayesian Learning and the Relevance Vector Machine},''
  \emph{Journal of Machine Learning Research}, vol.~1, no.~3, pp. 211--244,
  2001.

\bibitem{cslaplace}
S.~D. {Babacan}, R.~{Molina}, and A.~K. {Katsaggelos}, ``{Bayesian Compressive
  Sensing Using Laplace Priors},'' \emph{IEEE Transactions on Image
  Processing}, vol.~19, no.~1, pp. 53--63, 2010.

\bibitem{VAMP}
Winn, John, Bishop, M.~Christopher, Jaakkola, and Tommi, ``{Variational Message
  Passing},'' \emph{Journal of Machine Learning Research}, vol.~6, pp.
  661--694, 2005.

\bibitem{MF2002}
E.~P. Xing, M.~I. Jordan, and S.~Russell, ``{A Generalized Mean Field Algorithm
  for Variational Inference in Exponential Families},'' in \emph{Nineteenth
  Conference on Uncertainty in Artificial Intelligence}, 2002.

\bibitem{Reviewer3R1}
L.~{He}, Y.~{Wu}, S.~{Ma}, T.~{Ng}, and H.~V. {Poor}, ``{Superimposed
  Training-Based Channel Estimation and Data Detection for OFDM
  Amplify-and-Forward Cooperative Systems Under High Mobility},'' \emph{IEEE
  Transactions on Signal Processing}, vol.~60, no.~1, pp. 274--284, 2012.

\bibitem{Reviewer3R2}
K.~{Zhong}, Y.~{Wu}, and S.~{Li}, ``{Signal Detection for OFDM-Based Virtual
  MIMO Systems under Unknown Doubly Selective Channels, Multiple Interferences
  and Phase Noises},'' \emph{IEEE Transactions on Wireless Communications},
  vol.~12, no.~10, pp. 5309--5321, 2013.

\bibitem{GuoAConcise}
Q.~{Guo} and D.~D. {Huang}, ``{A Concise Representation for the Soft-in
  Soft-out {LMMSE} Detector},'' \emph{IEEE Communications Letters}, vol.~15,
  no.~5, pp. 566--568, 2011.

\bibitem{extended}
S.~J. {Julier} and J.~K. {Uhlmann}, ``Unscented filtering and nonlinear
  estimation,'' \emph{Proceedings of the IEEE}, vol.~92, no.~3, pp. 401--422,
  2004.

\bibitem{BIUTAMP}
Z.~{Yuan}, Q.~{Guo}, and M.~{Luo}, ``Approximate message passing with unitary
  transformation for robust bilinear recovery,'' \emph{IEEE Transactions on
  Signal Processing}, vol.~69, pp. 617--630, 2021.

\bibitem{1993Fundamentals}
S.~K. Sengijpta, ``{Fundamentals of Statistical Signal Processing: Estimation
  Theory},'' \emph{Technometrics}, vol.~37, no.~4, pp. 465--466, 1993.

\end{thebibliography}

\end{document}